%% file: example_paper.tex
\documentclass{article}

\usepackage{listings} 
\usepackage{xcolor} 

\lstdefinestyle{mypython}{
    language=Python,
    basicstyle=\ttfamily\footnotesize,
    keywordstyle=\color{blue},
    commentstyle=\color{gray},
    stringstyle=\color{red},
    numbers=left,
    numberstyle=\tiny\color{gray},
    stepnumber=1,
    breaklines=true,
    frame=single,
    captionpos=b
}

\usepackage{microtype}
\usepackage{graphicx}
\usepackage{subfigure}
\usepackage{booktabs} 

\usepackage{hyperref}

\usepackage[underreview]{icml2025}

\usepackage{amsmath}
\usepackage{amssymb}
\usepackage{mathtools}
\usepackage{amsthm}

\usepackage[capitalize,noabbrev]{cleveref}

\theoremstyle{plain}

\theoremstyle{definition}

\theoremstyle{remark}

\usepackage[textsize=tiny]{todonotes}

\icmltitlerunning{Koel-TTS}

\begin{document}

\twocolumn[
\icmltitle{Koel-TTS: Enhancing LLM based Speech Generation with Preference Alignment and Classifier Free Guidance}



\icmlsetsymbol{equal}{*}

\begin{icmlauthorlist}
\icmlauthor{Shehzeen Hussain}{equal,comp}
\icmlauthor{Paarth Neekhara}{equal,comp}
\icmlauthor{Xuesong Yang}{equal,comp}
\icmlauthor{Edresson Casanova}{comp}
\icmlauthor{Subhankar Ghosh}{comp}
\icmlauthor{Mikyas T. Desta}{comp}
\icmlauthor{Roy Fejgin}{comp}
\icmlauthor{Rafael Valle}{comp}
\icmlauthor{Jason Li}{comp}

\end{icmlauthorlist}

\icmlaffiliation{comp}{NVIDIA Corporation, Santa Clara, CA, USA}

\icmlcorrespondingauthor{Shehzeen Hussain}{shehzeenh@nvidia.com}
\icmlcorrespondingauthor{Paarth Neekhara}{pneekhara@nvidia.com}

\icmlkeywords{Machine Learning, ICML}

\vskip 0.3in
]



\printAffiliationsAndNotice{\icmlEqualContribution} 

\begin{abstract}
While autoregressive speech token generation models produce speech with remarkable variety and naturalness, their inherent lack of controllability often results in issues such as hallucinations and undesired vocalizations that do not conform to conditioning inputs. 
We introduce Koel-TTS, a suite of enhanced encoder-decoder Transformer TTS models that address these challenges by incorporating preference alignment techniques guided by automatic speech recognition and speaker verification models. 
Additionally, we incorporate classifier-free guidance to further improve synthesis adherence to the transcript and reference speaker audio. Our experiments demonstrate that these optimizations significantly enhance target speaker similarity, intelligibility, and naturalness of synthesized speech. 
Notably, Koel-TTS directly maps text and context audio to acoustic tokens, and on the aforementioned metrics, outperforms state-of-the-art TTS models, despite being trained on a significantly smaller dataset. Audio samples and demos are available on our website~\footnote{\url{https://koeltts.github.io/}}.

\end{abstract}

\section{Introduction}

\input{sections/introduction}


\section{Methodology}\label{sec:methodology}
\input{sections/methodology}

\section{Experiments}\label{sec:experiments}
\input{sections/experiments}
\section{Conclusion}
\input{sections/conclusion}

\bibliographystyle{icml2025}
\bibliography{references}\label{sec:references}

\newpage
\newpage
\appendix
\onecolumn
\input{appendix}

\end{document}

%% file: sections/introduction.tex

The advancement of large language models (LLMs) has brought transformative improvements to speech synthesis, enabling more natural and contextually adaptive speech generation. In particular, there has been a recent surge in the use of LLMs for various applications such as text-to-speech (TTS) and speech-to-speech translation~\cite{wang2023neural,zhang2023speak,borsos2023audiolm,t5tts,yanguniaudio,wang2024speechx}.  LLM-based TTS systems enable prompt-based customization, generating speech with human-like intonation while adapting to stylistic cues, contexts, and expressive nuances. This allows for diverse applications, from conversational interfaces to expressive narration, without extensive retraining. These advancements have been largely driven by the emergence of discrete neural audio codecs, which compress raw audio into token-based representations while preserving high fidelity~\cite{encodec,dac_kumar2024high,zeghidour2021soundstream,langman2024spectral,casanova2024low}.

Despite these advances, LLM-based TTS systems face challenges, with hallucinations being a prominent issue~\cite{sahoo2024comprehensive,song2024ella,t5tts,borsos2023audiolm}.
For example, when encountering text with repeated or redundant phrases, LLM-based TTS models may overemphasize these repetitions or fail to capture the intended flow and naturalness of the sentence. Additionally, among the multiple outputs sampled for the same input, there can be significant variation in quality, with some outputs sounding more natural, accurate, and appealing than others. This issue is akin to challenges faced in text-generation LLMs, where outputs may range from highly coherent to erroneous, depending on the model's response to complex prompts.

For text-generation, preference alignment techniques~\cite{christiano2017deep,ouyang2022rlhf,shao2024deepseekmath,rafailov2024direct, adler2024nemotron} have been proposed to guide models to produce outputs that better match human preferences in coherence, relevance, and clarity. This is achieved through training with human feedback or automated scoring, based on criteria such as factual correctness and fluency. Driven by these advances, recent research employs preference alignment algorithms, including RLHF~\cite{ouyang2022rlhf} and offline preference ranking methods~\cite{rafailov2024direct,azar2024ipo}, to refine audio LLM outputs. For instance, SpeechAlign~\cite{zhang2024speechalign}, proposes an iterative strategy to align speech language models with human preferences by addressing the distribution gap between golden AR tokens (from real speech) and synthetic AR tokens (generated during inference).  Although real speech from ground truth can be used to guide training, we will show that it introduces inconsistencies due to its fundamentally different distribution from model-generated tokens. This issue makes preference-based optimization such as DPO~\cite{rafailov2024direct} less effective. 
Nonetheless, this approach has been applied in scenarios where obtaining high-quality positive examples is particularly challenging \cite{chen2024self,zhang2024speechalign}. 

Another research direction to amplify the influence of conditioning inputs in generative models is Classifier-Free Guidance (CFG). 
CFG was originally introduced to trade-off sample fidelity and diversity without relying on a separate classifier in diffusion models~\cite{ho2021classifier}.  
Recently, CFG has been successfully explored in LLM-based text-generation models \cite{sanchez2023stay, fonseca2024can, smirnov2024classifier}. In the context of text-to-speech synthesis, CFG has been extended to improve non-autoregressive flow-matching models (CFM)~\cite{le2024voicebox, du2024cosyvoice, chen2024f5, eskimez2024e2}. However, the applicability of CFG for enhancing LLM-based speech token prediction models is underexplored, with only a few attempts at improving textual coherence~\cite{darefsky2024parakeet}.

Building upon the above insights, 
we propose preference alignment and CFG techniques to enhance contextual coherence of LLM-based TTS models.
We introduce Koel-TTS, a transformer-based TTS model that leverages a low-frame-rate ($21.5$ FPS) audio codec~\cite{casanova2024low} to enable low-latency autoregressive speech generation. 
To perform preference alignment, we first identify key metrics that strongly correlate with human judgments of generated speech: transcription accuracy and target speaker similarity. Each metric captures distinct aspects of the generated output and can be evaluated using automatic speech recognition (ASR) and speaker verification (SV) models. We integrate these metrics into a reward system that ranks the generated outputs. 
With this foundation, we then explore preference alignment algorithms, focusing on pairwise ranking methods and scalar reward optimization. 
Our findings show that fine-tuning the base model with preference alignment significantly improves speaker similarity, intelligibility, and generalization to unseen speakers. Notably, our method also enhances naturalness, despite not explicitly optimizing for this metric.

To further enhance synthesis quality with CFG, we train the Koel-TTS model with both conditional and unconditional inputs, by randomly dropping out conditional information (text and context audio) during training.
During inference, the unconditional logits are combined with conditional logits using a CFG scale, to achieve significant improvement in intelligibility, speaker similarity, and naturalness of generated speech. Furthermore CFG can be applied independently to the base model or the preference aligned model, yielding substantial improvements across all
evaluation metrics for both. Combining preference alignment with CFG, we train a 1.1 billion parameter multilingual Koel-TTS model that
achieves state-of-the-art zero-shot TTS results across several human and automatic evaluation metrics.

The key contributions of this work are as follows:
\begin{itemize}
\vspace{-0.3cm}
\item We introduce Koel-TTS, a multilingual encoder-decoder transformer model that maps text and context audio directly to acoustic tokens using a low-frame-rate audio codec, enabling expressive, robust, and low-latency autoregressive speech synthesis.
\vspace{-0.2cm}
\item We propose a novel preference alignment framework for LLM-based TTS by leveraging ASR and SV models as reward signals, significantly improving speaker similarity, intelligibility, and generalization to seen and unseen speakers.
\vspace{-0.2cm}
\item We adapt classifier-free guidance, dropping both text and context conditioning, to enhance LLM-based speech synthesis, demonstrating its effectiveness in improving naturalness, speaker similarity, and intelligibility.
\vspace{-0.2cm}
\item Our Koel-TTS model, trained with preference alignment and CFG, achieves state-of-the-art zero-shot TTS performance while reducing hallucinations and improving intelligibility. Our model implementation is publicly available in the Koel-TTS repository\footnote{Omitted from blind review}.
\vspace{-0.2cm}

\end{itemize}

\begin{figure*}[tp]
    \centering
    \includegraphics[width=0.9\textwidth]{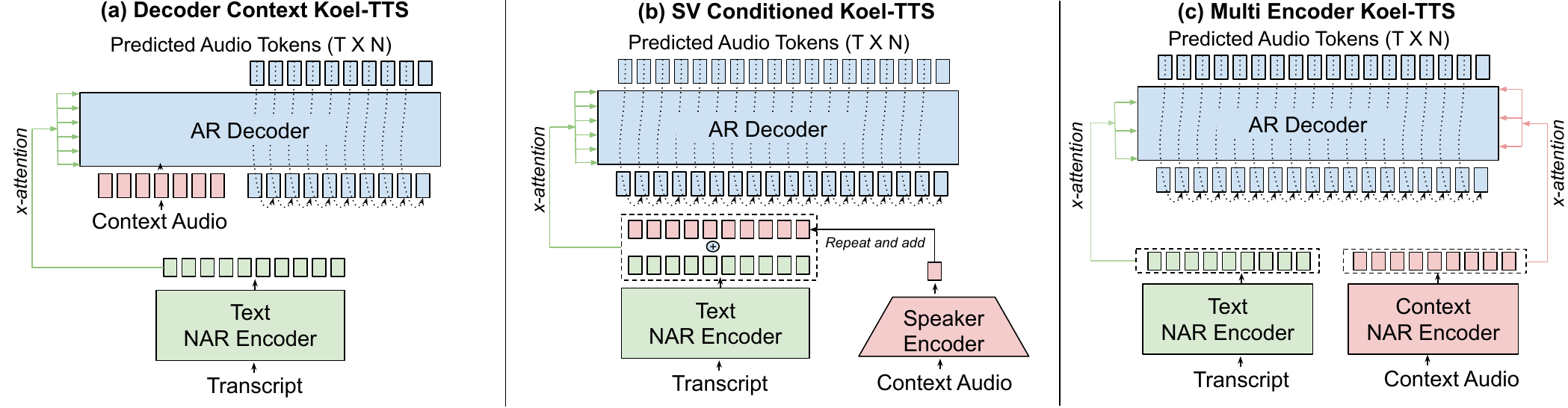}
\vspace{-2mm}    
\caption{
Koel-TTS Model Architectures: Three methods for conditioning TTS synthesis on context audio and transcripts. The \textit{Decoder Context} approach utilizes the decoder's self-attention mechanism for speaker conditioning. The \textit{Multi Encoder} and \textit{SV Conditioned} models employ cross-attention layers for speaker conditioning.
}
\vspace{-4mm}
    \label{figs:model_overview}
\end{figure*}

%% file: sections/methodology.tex
Our proposed framework is an autoregressive speech token generation model that is conditioned on a text input and a context audio.
In this section, 
we first
describe the tokenization scheme employed for representing speech and text. Next, we detail the model architecture and describe three model designs we explore for context-conditioning.
Finally, we propose two key techniques to improve the robustness and speaker similarity of our model using preference optimization algorithms and classifier free guidance.

\subsection{Tokenization}
\textbf{Speech:} We employ a neural audio codec model to transform raw speech signals into tokenized representations. For a given audio signal $\mathbf{a}$, the codec model outputs a two-dimensional acoustic matrix $\mathbf{C}_{T \times N}=\textit{CodecModel}(\mathbf{a})$. In this representation, $\mathbf{C}_{T \times N}$ consists of $m$-bit discrete codes, where $T$ corresponds to the downsampled sequence length, and $N$ represents the number of codebooks per timestep. We utilize the Low Frame-rate Speech Codec \cite{casanova2024low}, which achieves high-quality audio compression at a bitrate of $1.89$ kbps and a frame rate of $21.5$ frames per second, utilizing $N$\texttt{=8} independent codebooks.  
The codec uses  Finite Scalar Quantization (FSQ) \cite{mentzer2024finite}, which ensures independence among the codebooks. 
This independence eliminates the need for additional models or delay mechanisms, enabling the parallel prediction of all $N$ codebooks at each timestep.

\textbf{Text:} 
We explore two text tokenization methods: phonemes and characters. Phonemes, commonly used in neural TTS, capture fundamental sound units but require language-specific grapheme-to-phoneme (G2P) conversion. In contrast, character-based tokenization eliminates this need, enabling direct conversion of graphemes to acoustic information. In our experiments, we use IPA phonemes and character tokenizers for English, German, and Spanish, while applying only character tokenizers for other languages.
We utilize an aggregated tokenizer that maintains separate token embeddings for each language. Additionally, we perform an ablation study with a shared character tokenizer and a multilingual sentencepiece tokenizer across all languages, with results detailed in Appendix~\ref{sec:multilingualablations}.


\subsection{Model Overview}

Our speech generation model comprises an autoregressive (AR) transformer decoder conditioned on text encodings from a non-autoregressive (NAR) transformer encoder, using cross-attention layers. 
The AR transformer predicts audio tokens frame by frame, generating all $N$ codebooks in parallel at each time step, conditioned on previous generations and inputs to the cross-attention layers.
To influence the speaker and style of the generated audio through context audio conditioning, we explore three model designs as illustrated in Figure~\ref{figs:model_overview},

\textbf{SV Conditioned Koel-TTS:} In this configuration, a speaker embedding vector is extracted from the context audio using a pre-trained SV model~\cite{koluguri2022titanetlarge}. This embedding vector is projected to hidden dimension of the transformer network, temporally expanded (repeated across the time axis) and added to the text encoder's output. 
The resulting combined representation serves as the input to the cross-attention layers of the AR decoder, enabling the prediction of audio codes while conditioning on the speaker identity. 
The advantage of this design is the ability to leverage transfer learning from the SV model, thereby enhancing generalization in scenarios with limited data. 
However, since the speaker vector is a compressed representation that primarily preserves voice identity, it does not capture other nuanced aspects of the context audio, such as speaking style and accent, which limits control over the generated speech.

\textbf{Decoder Context Koel-TTS:} Here, the context audio tokens are directly provided as input to the AR decoder by prepending them to the target audio tokens. This approach eliminates the need for a separate audio encoder to process context audio. Instead, a single unified transformer decoder processes both the context and target audio tokens, leveraging a shared representation for conditioning and prediction.

\textbf{Multi encoder Koel-TTS:} In this architecture, context audio tokens are processed by a dedicated context encoder, which is a separate NAR transformer encoder. The outputs of the context encoder and text encoder are fed into alternate cross-attention layers of the AR decoder, as illustrated in Figure~\ref{figs:model_overview}c. This design allows for a clear separation of modalities, where each encoder operates independently, and the decoder employs dedicated cross-attention mechanisms to integrate the outputs.
This model also allows cross-attention biasing over the text tokens independently for learning monotonic alignment, while allowing variable length context audios.

\subsection{Training Objective}
The output of each decoder-timestep is mapped to a vector of size $N \times 2^m$ using a linear layer to obtain the logits of all $N$ codebooks (each of size $m$-bits) at that timestep. Thereby, for all decoder time-steps, we obtain logits $\ell$ of size $T \times N \times 2^m$ and calculate cross-entropy as follows:
\vspace{-2mm}
\begin{align*}
\mathcal{L}_\textit{token} = \textit{CE}\left(\textit{softmax}\left(\ell\right), \textit{target}_{N\times T}\right)
\end{align*}
In addition to the above, to improve text and speech alignment, past work~\cite{t5tts} recommends biasing the cross-attention scores between the transcript encoder and AR decoder to be monotonic, using an attention prior and Connectionist Temporal Classification (CTC) loss. Specifically, given the cross-attention score matrix $\mathbf{A}_{T\times M}^{\textit{l,h}}$,  of the $h^{\textit{th}}$ cross-attention head in decoder layer \textit{l}, between the audio timesteps ($T$) and text timesteps ($M$), we generate a static prior using the 2D beta-binomial distribution $\mathbf{P}_{T \times M}$.
Given this prior, we obtain the re-scaled attention scores as: 
$$\mathbf{A}_{T\times M}^{\textit{l,h}} \gets \mathbf{A}_{T\times M}^{\textit{l,h}} \odot  \mathbf{P}_{T \times M}$$ 
The attention prior is applied for the first $10{,}000$ training iterations and then linearly annealed to a uniform distribution (all ones) for the next $5{,}000$ iterations and turned off thereafter. Turning off the prior is necessary since we cannot use this prior during inference\footnote{The final audio sequence length is unknown during inference.} and annealing ensures stability during training.

Additionally, to encourage valid monotonic sampling from the alignment matrix, we calculate likelihood of all possible monotonic reductions using the CTC algorithm. That is, given the alignment matrix $\mathbf{A}^{\textit{soft}_{l,h}}_{T\times M}=\textit{softmax}(\mathbf{A}_{T\times M}^{\textit{l,h}})$, we obtain the alignment loss for a decoder layer and head as: $$\mathcal{L}_{\textit{align}}^{l,h} = \textit{CTCLoss}\left(\mathbf{A}^{\textit{soft}_{l,h}}_{T\times M}, q_{M}\right)$$
where $q_{M}=\{1, 2, \dots M\}$ is the target monotonic sequence from $1$ to $M$. The alignment loss is summed across all cross-attention heads and layers to obtain $\mathcal{L}_{\textit{align}} = \sum_{\substack{l,h}} \mathcal{L}_{\textit{align}}^{l,h}$.
The final training loss is obtained as $\mathcal{L} = \mathcal{L}_\textit{token} + \alpha \mathcal{L}_{\textit{align}}$, where $\alpha$ is the alignment loss coefficient.

\begin{figure*}[tp]
    \centering
    \includegraphics[width=1.0\textwidth]{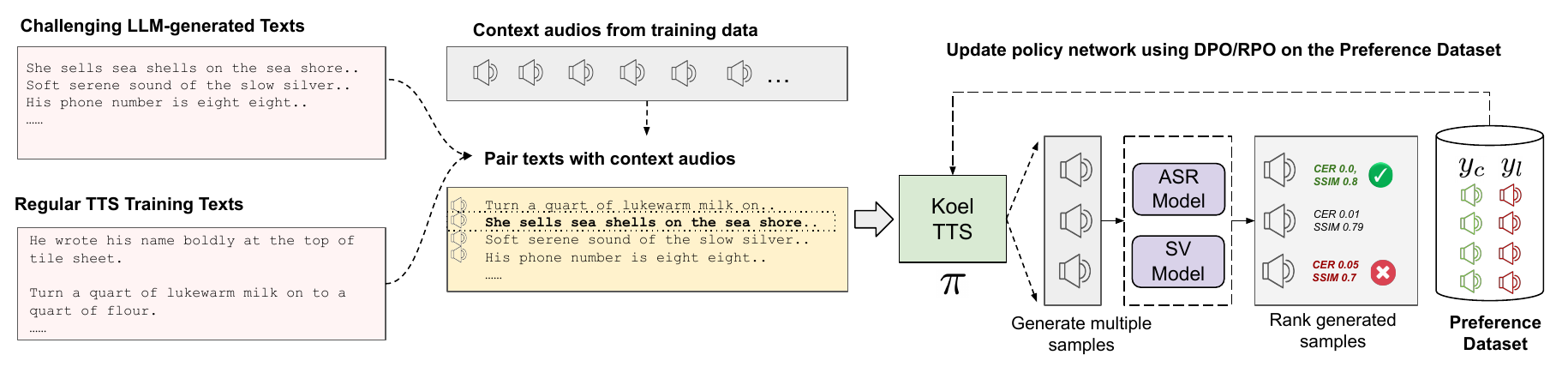}
\vspace{-8mm}    
\caption{Preference Alignment for Koel-TTS: Koel-TTS generates multiple outputs for challenging text and context audio prompts, which are then rewarded using ASR and SV Models to create a preference dataset for DPO and RPO.}
\vspace{-4mm}
\label{figs:dpo_diag}
\end{figure*}

\subsection{Preference Alignment}\label{sec:pref_align}

We employ preference optimization methods to steer the outputs of Koel-TTS towards more desirable results. 
For a given text and context audio input $x=(x_\textit{text}, x_\textit{audio})$, the model's response distribution $\pi(y|x)$ encompasses a range of potential outputs $y$ with varying levels of alignment to the desired criteria. By constructing a dataset that explicitly labels certain responses $y_c$ as chosen and others $y_l$ as rejected, we can leverage preference-based optimization algorithms to shift the model's distribution toward producing more preferred responses.

One such approach is Direct Preference Optimization (DPO)~\cite{rafailov2024direct}. DPO uses preference comparisons to modify the policy $\pi$ by contrasting it against a reference policy $\pi_{\text{ref}}$. Specifically, given an input $x$ and a chosen response $y_c$ that is preferred over a rejected response $y_l$, DPO seeks to increase the likelihood ratio $\frac{\pi(y_c|x)}{\pi_{\text{ref}}(y_c|x)}$ relative to $\frac{\pi(y_l|x)}{\pi_{\text{ref}}(y_l|x)}$. The core objective can be expressed as:
\vspace{-2mm}
\begin{align*}
\mathcal{L}_{\text{DPO}} = \mathbb{E}_{x, y_c, y_l}\left[ \beta \log\frac{\pi(y_c|x)}{\pi_{\text{ref}}(y_c|x)} - \beta \log\frac{\pi(y_l|x)}{\pi_{\text{ref}}(y_l|x)} \right].
\end{align*}

where $\beta$ is a parameter for controlling the deviation from the base reference policy $\pi_{\text{ref}}$. The above formulation, encourages $\pi$ to produce responses more similar to $y_c$ than $y_l$, effectively aligning the model with the desired preferences.

Building upon DPO, we also leverage Reward-aware Preference Optimization (RPO)~\cite{adler2024nemotron}, which considers the magnitude of reward differences in the optimization process.
Rather than treating all chosen versus rejected distinctions as equal, RPO utilizes scalar rewards to measure how much better the chosen response is compared to the rejected one. The RPO objective introduces a factor that scales the preference updates based on the reward gap $r^*(x,y_c) - r^*(x,y_l)$ as follows:

\vspace{-4mm}
\begin{align*}
\mathcal{L}_{\text{RPO}}(x, y_c, y_l) 
= \mathbb{D} \Bigg[ 
    & \beta \log \frac{\pi(y_c \mid x)}{\pi_{\text{ref}}(y_c \mid x)} 
    - \beta \log \frac{\pi(y_l \mid x)}{\pi_{\text{ref}}(y_l \mid x)} \\
    & \;\Big\|\; 
    \eta \big(r^\ast(x, y_c) - r^\ast(x, y_l)\big)
\Bigg],
\end{align*}

where \(\eta\) is a scaling parameter and $\mathbb{D}$ is a distance metric given by 
$\mathbb{D}\left[a\|b\right] := \sigma(b) \log \frac{\sigma(b)}{\sigma(a)} + (1-\sigma(b)) \log \frac{1-\sigma(b)}{1-\sigma(a)}$
Thereby, RPO mitigates overfitting to narrowly better responses since the loss value is scaled as per the reward difference.

\textbf{Preference Data Creation and Reward System:}
To construct the preference dataset $(x, y_c, y_l)$, we begin by selecting a diverse set of text and speaker prompt combinations that challenge the model's ability to produce accurate and natural speech. The text data includes a mix of regular sentences from standard TTS datasets, and carefully curated challenging transcripts generated by prompting text LLMs. These challenging texts are designed to test the model's robustness and include elements such as repeated words, numbers, and phonetically complex sequences. The inclusion of standard texts ensures generalizability, while the challenging examples target specific weaknesses of the TTS model as illustrated in Figure~\ref{figs:dpo_diag}. 

For each text and speaker prompt, we generate $P$ audio samples using multinomial top-k sampling ($k$\texttt{=80}, temperature=\texttt{0.7}). 
Each generation is evaluated using the Parakeet TDT 1.1B ASR~\cite{xu2023efficient} and Titanet-large SV~\cite{koluguri2022titanet} models. Specifically, we obtain the character error rate (CER) between the transcript of the generated audio and input text using the ASR model, and cosine similarity (SSIM) between the embeddings of the context audio and the generated audio obtained from the SV model. 
Based on the CER and SSIM, we perform Pareto optimal ranking~\cite{deb2011multi} on the set of $P$ generated audio samples for a given input pair---First, we identify the Pareto front, which consists of all audio samples that are not dominated by any other sample. That is, no other audio is strictly better on at least one metric and equally good or better on the other. Once we identify the first Pareto front, we remove those samples and repeat the process on the remaining audios to find the next front, and so on.
Within each Pareto front, we prioritize samples by assigning higher ranks to those with lower CER scores. If there are ties based on CER, we further differentiate by favoring samples with higher SSIM values. We detail this procedure in Appendix~\ref{paretorankingalgo}.

After ranking the examples, we select the highest-ranked as chosen and the lowest-ranked as rejected for DPO, since we empirically find high-contrast pairs to be beneficial for DPO. For RPO, which handles scalar reward differences, we pair the top two with the bottom two in all combinations. In both cases, we discard pairs where the chosen example scores worse on any metric (CER or SSIM) than the rejected one.

To assign scalar rewards for RPO, we normalize the CER and SSIM differences between the chosen and rejected examples, and set the reward gap as: 
$$r^*(x,y_c) - r^*(x,y_l) = \Phi (\tilde{\Delta \text{CER}}) + \Phi (\tilde{\Delta \text{SSIM}})$$
where $\Phi$ is the cumulative distribution function (CDF) of the standard normal distribution, and $\tilde{\Delta \text{CER}}$ and $\tilde{\Delta \text{SSIM}}$ are the normalized differences of CER and SSIM respectively, between the chosen and rejected examples.


\subsection{Classifier Free Guidance}\label{sec:cfg}
\input{sections/cfg}

%% file: sections/cfg.tex
To adapt CFG for autoregressive token prediction models, we train both a conditional and an unconditional model simultaneously by randomly dropping out the text and context/speaker conditioning during training. At inference time, conditional and unconditional outputs are combined to guide the speech generation process. This approach allows for more precise control over the generated speech, which can lead to improved pronunciation, prosody, robustness, and overall audio quality. 

\begin{figure}[!ht]
    \centering
    \includegraphics[width=1.0\columnwidth]{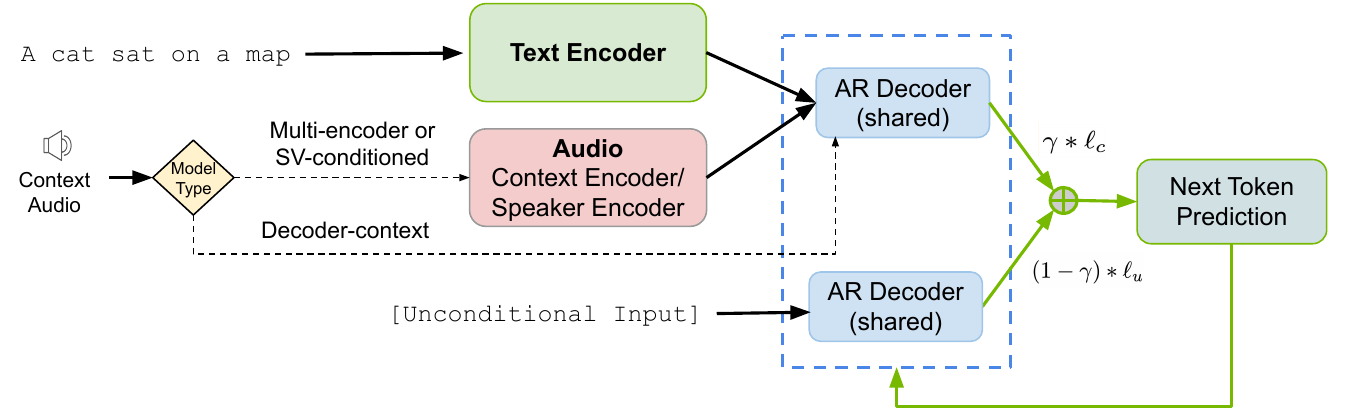}
    \vspace{-8mm} 
    \caption{\footnotesize{CFG Inference: Logits from conditional and unconditional inputs are combined using a CFG scale $\gamma$\texttt{>1}, steering model predictions towards better alignment with conditional inputs.}}
    \label{figs:cfg}
     \vspace{-4mm} 
\end{figure}

Distinct from the previous work that only deals with text-independent conditionals~\cite{darefsky2024parakeet}, in our approach, we randomly dropout both audio and text conditioning inputs (with a probability of $10\%$) during training and interpolate conditional logits ($\ell_{c}$) with the unconditional logits ($\ell_{u}$) during inference\footnote{Note that CFG inference doubles the effective batch size to obtain conditional and unconditional logits.}, 
\begin{equation*}
    \ell_\textit{cfg} = \gamma * \ell_{c} + (1 - \gamma) * \ell_{u}
\end{equation*}
where $\gamma\ge1$ is the CFG interpolation scale controlling the strength of guidance. Higher scale values steer the generation to follow the text/audio inputs, while lower scale values allow more variations. In practice, we sweep around a range of values to find the optimal scale $\gamma$. 
Figure~\ref{figs:cfg} demonstrates the CFG inference process. 

%% file: sections/experiments.tex
\subsection{Datasets, Training and Evaluation Criteria}
\label{sec:datasets}

For our primary experiments, we train the models on a data-blend containing $18k$ hours of English TTS data from the following datasets: \emph{train-clean-360} and \emph{train-clean-100} subsets of LibriTTS~\cite{zen2019libritts}, HiFiTTS~\cite{bakhturina21_interspeech}, a $17k$-hour subset of the LibriVox MLS dataset~\cite{pratap20_interspeech} and a proprietary, 2-speaker, 63-hour dataset. 
With this dataset we create \textit{(context audio, transcript, target audio)} triplets where context and target audio are distinct utterances from the same speaker. During training, we use a random $5$ second context slice for the decoder context model and $3$ to $8$ seconds context slice for \textit{multi encoder} and \textit{SV conditioned} models. Model architectures and training details are provided in Appendix~\ref{sec:trainingdetails}.


\textbf{Evaluation:} We evaluate synthesized speech on intelligibility, speaker similarity, and naturalness. Intelligibility is measured using ASR-based character error rate (CER) and word error rate (WER), with Parakeet-TDT ~\cite{xu2023efficient} for English and \textit{whisper-large-v3}~\cite{radford2022whisper} for other languages. Speaker similarity is assessed via cosine similarity between speaker embeddings of synthesized and context audio, using \textit{Titanet-Small}~\cite{koluguri2022titanet} for embedding extraction—different from \textit{Titanet-Large}~\cite{koluguri2022titanetlarge} used in \textit{SV conditioned} model training. Naturalness is evaluated with Squim-MOS~\cite{kumar2023torchaudio}, and we also conduct a human evaluation for two of our zero-shot Koel-TTS models, benchmarking them against others in Section~\ref{sec:pastworkcomparison}.
For inference, we use multinomial top-k sampling ($k$\texttt{=80}, temperature\texttt{=0.6}). Due to probabilistic generation, each experiment is repeated five times, reporting mean metrics with 95\% confidence intervals.

For seen speaker TTS evaluation, we consider a test set of $200$ held-out utterances from \textit{train-clean-360} LibriTTS dataset. For unseen speakers, we create a subset of \textit{test-clean} LibriTTS containing $180$ utterances from a total of $36$ out of the $40$ speakers, using $5$ distinct context and target audios from each each speaker. We use a random $5$ second slice from the context audio during inference for all experiments.

\subsection{Architecture Comparison}
Table~\ref{tab:architecture_comp} presents the baseline results of different Koel-TTS architectures on seen and unseen English speakers, \textbf{without} incorporating preference alignment training or CFG inference.  
All three architectures achieve similar intelligibility, but the \textit{decoder context} model outperforms the \textit{multi encoder} model on unseen speaker similarity, while the latter performs slightly better on seen speakers.  
These results suggest that \emph{decoder context} model generalizes better to unseen speakers making it a more suitable choice for zero-shot TTS. 
The \textit{multi encoder} architecture tends to overfit to the training speakers, as indicated by consistently worse speaker similarity on unseen speakers, and better speaker similarity on seen speakers across all our experiments. 
While \textit{SV conditioned} model also achieves similar SSIM as decoder context, perceptually, we find the decoder context model captures the intended style of the context audio better. We encourage readers to listen to audio examples on our website.

\setlength{\tabcolsep}{2pt}
\begin{table}[ht]
\vspace{-4mm}
\caption{\footnotesize{Baseline TTS results on seen and unseen speakers for different Koel-TTS models, \textbf{without using CFG or preference alignment}. Lower CER(\%) \& WER(\%) indicate higher intelligibility. Higher SSIM indicates higher speaker similarity to ground-truth.}}
\centering
\resizebox{\columnwidth}{!}{%
\begin{tabular}{c|l|cccc}
\toprule
Eval Set & Model & CER(\%) $\downarrow$ & WER(\%) $\downarrow$ & SSIM $\uparrow$ & Squim-MOS $\uparrow$ \\
\midrule
& Ground Truth & $0.51 \pm 0.00$& $1.42 \pm 0.00 $  & $0.763 \pm 0.000$ & $4.616 \pm 0.03$\\
 Seen &  Decoder context & $1.73 \pm 0.60$ & $2.98 \pm 0.59$ & $0.700 \pm 0.001$ & $4.350 \pm 0.038$ \\ 
Speakers & SV Conditioned & $\mathbf{1.71 \pm 0.41}$ & $\mathbf{2.82 \pm 0.41}$ & $0.697 \pm 0.003$ & $\mathbf{4.360 \pm 0.021}$ \\ 
  & Multi Encoder  & $1.92 \pm 0.68$ & $3.02 \pm 0.76$ & $\mathbf{0.712 \pm 0.002}$ & $4.346 \pm 0.028$ \\ 
\midrule
 & Ground Truth & $0.80 \pm 0.00 $ & $1.83 \pm 0.00 $ & $0.771 \pm 0.000$ & $4.588 \pm 0.020$ \\
Unseen & Decoder  Context & $2.68 \pm 1.13$ & $4.02 \pm 1.12$ & $\mathbf{0.637 \pm 0.008}$ & $\mathbf{4.347 \pm 0.024}$ \\ 
Speakers& SV Conditioned & $3.12 \pm 0.98$ & $4.22 \pm 1.02$ & $0.619 \pm 0.003$ & $4.318 \pm 0.034$ \\ 
& Multi Encoder  & $\mathbf{2.56 \pm 1.44}$ & $\mathbf{3.74 \pm 1.36}$ & $0.601 \pm 0.004$ & $4.318 \pm 0.059$ \\
\bottomrule
\end{tabular} 
}
\label{tab:architecture_comp}
\vspace{-2mm}
\end{table}

\subsection{Preference Alignment}
To perform preference alignment, we create a preference dataset using the procedure described in Section~\ref{sec:pref_align}. 
Specifically, we first curate $800$ challenging texts generated using Llama-8b~\cite{touvron2023llama}. It is prompted to generate texts with repeated words and alliterations. The complete list of these texts and the prompts used for generating them can be found on our webpage. We pair each challenging text with $10$ random context audios sampled from our training dataset. Next, we sample $50{,}000$ regular transcripts from our training data, and pair each text with one random context audio from our training data. This results in a total $58{,}000$ text and context audio pairs. For each pair, we generate $6$ audio samples from each of our models and create chosen-rejected pairs using the reward and filtering criteria outlined in Section~\ref{sec:pref_align}.


\setlength{\tabcolsep}{3pt}
\begin{table*}[t]
\vspace{-2mm}
\caption{\footnotesize{Preference Alignment (DPO, RPO) and CFG evaluations for a multi encoder baseline model on seen and unseen speakers. Both methods improve intelligibility, speaker similarity and naturalness metrics, with best results achieved when they are used together. GT as Chosen indicates the ablation experiment considering ground truth audio as the chosen audio in the preference pairs.}}
\centering
\resizebox{\textwidth}{!}{%
\begin{tabular}{l|cccc|cccc}
\multicolumn{1}{c}{} & \multicolumn{4}{c}{\emph{Seen Speakers}} & \multicolumn{4}{c}{\emph{Unseen Speakers}} \\
\toprule
Model/Technique & CER(\%) $\downarrow$ & WER(\%) $\downarrow$ & SSIM $\uparrow$ & Squim-MOS $\uparrow$ & CER(\%) $\downarrow$ & WER(\%) $\downarrow$ & SSIM $\uparrow$ & Squim-MOS $\uparrow$ \\
\midrule
Ground Truth & $0.51 \pm 0.00$& $1.42 \pm 0.00 $  &  $0.763 \pm 0.000$ & $4.616 \pm 0.03$ & $0.80 \pm 0.00 $ & $1.83 \pm 0.00 $ & $0.771 \pm 0.000$ & $4.588 \pm 0.020$  \\
\midrule
Multi Encoder (BL) & $1.92 \pm 0.68$ & $3.02 \pm 0.76$ & $0.712 \pm 0.002$ & $4.346 \pm 0.028$ & $2.56 \pm 1.44$ & $3.74 \pm 1.36$ & $0.601 \pm 0.004$ & $4.318 \pm 0.059$ \\ 
BL + RPO & $1.01 \pm 0.58$ & $1.76 \pm 0.59$ & $0.737 \pm 0.002$ & $4.408 \pm 0.010$ & $0.79 \pm 0.12$ & $1.72 \pm 0.18$ & $0.641 \pm 0.002$ & $4.389 \pm 0.021$ \\ 
BL + DPO & $0.67 \pm 0.17$ & $1.48 \pm 0.34$ & $0.737 \pm 0.004$ & $4.406 \pm 0.011$ & $0.62 \pm 0.12$ & $1.49 \pm 0.20$ & $0.645 \pm 0.001$ & $4.402 \pm 0.029$ \\  
BL + DPO (GT as Chosen) & $1.58 \pm 0.42$ & $2.88 \pm 0.41$ & $0.710 \pm 0.005$ & $4.344 \pm 0.038$ & $2.67 \pm 0.006$ & $4.08 \pm 0.48$ & $0.522 \pm 0.004$ &  $4.281 \pm 0.055$
\\
\midrule
BL + CFG ($\gamma=2.5$) & $0.73 \pm 0.16$ & $1.63 \pm 0.24$ & $0.752 \pm 0.003$ & $4.420 \pm 0.010$ & $0.69 \pm 0.07$ & $1.59 \pm 0.10$ & $0.653 \pm 0.005$ & $4.415 \pm 0.007$ \\ 
BL + RPO + CFG ($\gamma=2.5$)& $0.75 \pm 0.23$ & $1.56 \pm 0.32$ & $0.766 \pm 0.003$ & $\mathbf{4.422 \pm 0.011}$ & $\mathbf{0.51 \pm 0.12}$ & $\mathbf{1.25 \pm 0.18}$ & $0.674 \pm 0.004$ & $4.392 \pm 0.029$ \\ 
BL + DPO + CFG ($\gamma=2.5$)& $\mathbf{0.51 \pm 0.12}$ & $\mathbf{1.32 \pm 0.23}$ & $\mathbf{0.767 \pm 0.004}$ & $4.418 \pm 0.012$ & $0.58 \pm 0.17$ & $1.38 \pm 0.08$ & $\mathbf{0.678 \pm 0.005}$ & $\mathbf{4.417 \pm 0.015}$ \\ 
\bottomrule
\end{tabular} 
}
\label{tab:dpo_rpo_cfg_base_multi}
\vspace{-4mm}
\end{table*}

Starting from our base checkpoints, we perform DPO or RPO finetuning for a maximum of $4{,}000$ mini-batch iterations using a batch-size of $64$ pairs, optimized using Adam optimizer with a fixed learning rate (LR) \texttt{2e-7}. We set $\beta$\texttt{=0.01} and $\eta$\texttt{=1.0} as the default loss coeffecients in DPO and RPO. We choose the checkpoint with the lowest validation loss and evaluate the preference-aligned models for intelligibility, speaker similarity  and naturalness. Table~\ref{tab:dpo_rpo_cfg_base_multi} shows the results of these experiments on the \textit{multi encoder} model. Results of DPO/RPO across all architectures can be found in Appendix~\ref{sec:dporpoall}.
As evident, both DPO and RPO significantly improve intelligibility and speaker-similarity metrics over the base model across all architectures. Interestingly, preference alignment also significantly improves zero-shot speaker similarity on unseen speakers, even though we do not include any new speakers in the preference data creation. The naturalness metric Squim-MOS also shows an improvement over the baseline models, even though we don't explicitly include it in our reward system. 
This suggests that CER and SSIM metrics serve as good proxy for human preferences and can be automatically computed, thereby allowing easy scaling up of the preference alignment process.
We find RPO to be less sensitive to hyper-parameter tuning and number of training iterations than DPO. 
RPO also works reliably when preference data does not have high-contrast chosen-rejected pairs, since it considers reward differences instead of a binary pair, in its optimization objective.

To compare with a prior technique proposed in SpeechAlign ~\cite{zhang2024speechalign}, we perform an ablation in which the ground-truth audios are selected as our chosen examples (as opposed to generated samples).  
We use a subset of \textit{train-clean-360} LibriTTS data as chosen examples, and the worst ranked of the $6$ generations (for an input) as the rejected example, creating a similar sized preference dataset as our other experiments.
We find that preference alignment algorithms find it trivial to differentiate ground-truth examples from generated ones with the preference loss reducing to nearly zero within the first few hundred iterations.
With our default DPO hyperparameters ($\beta$\texttt{=0.01}, LR\texttt{=2e-7}) such a setup leads to model degeneration and a very high CER (\texttt{>90\%}). 
Fine-tuning DPO hyperparameters ($\beta$\texttt{=1.0}, LR\texttt{=1e-7}) and early stopping, 
prevents model degeneration, but does not yield improvement over the baseline model (Table~\ref{tab:dpo_rpo_cfg_base_multi}). This suggests it is important to obtain chosen-rejected pairs from model's output distribution, for preference optimization such as DPO to work effectively.

\subsection{Classifier Free Guidance}
\begin{figure}[!ht]
    \centering
    \includegraphics[width=1.0\columnwidth]{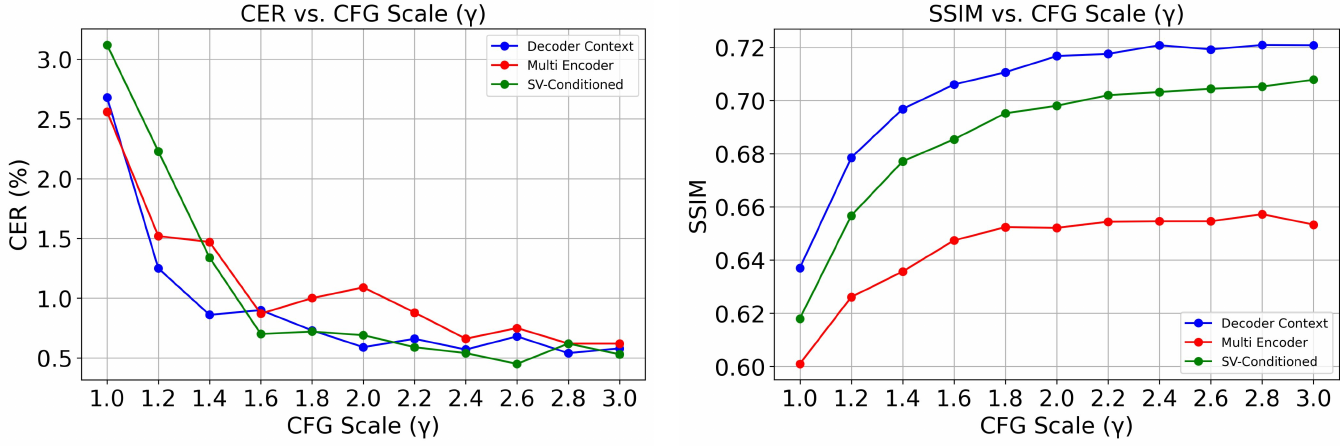}
    \vspace{-8mm} 
    \caption{Effect of CFG scale on CER/SSIM for unseen speakers}
    \label{figs:cfgplots}
\end{figure}

By controlling the CFG scale $\gamma$ during inference, we can steer the generations to be better aligned with conditional inputs.
We vary $\gamma$ between $1$ to $3$ at $0.2$ intervals and show the results of this experiment on unseen speakers in Figure~\ref{figs:cfgplots}. 
Increasing $\gamma$ significantly reduces the CER and simultaneously increases SSIM across all models. 
From these observations, we set $\gamma$\texttt{=2.5} as the optimal value. 

\vspace{-2mm} 

\begin{figure*}[tp]
    \centering
    \includegraphics[width=0.97\textwidth]{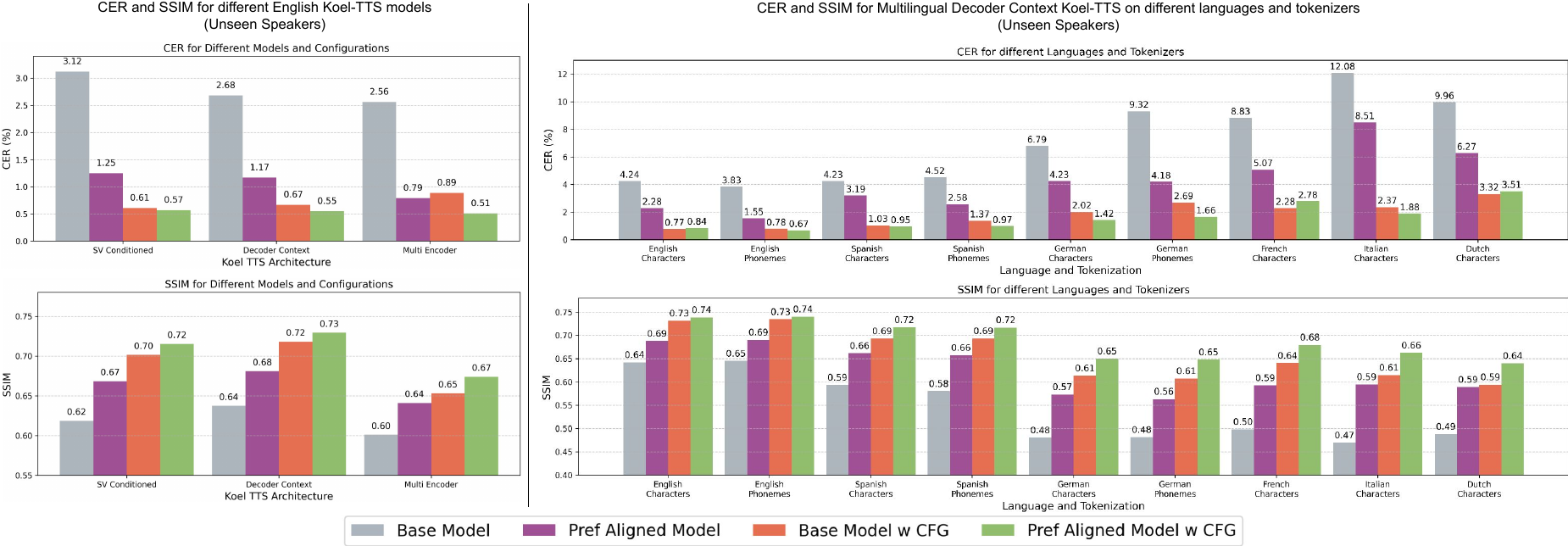}
\vspace{-4mm}    
    \caption{Performance of Koel-TTS Architectures. \textbf{Left:} 
    Intelligibility (CER) and speaker similarity (SSIM) evaluations of base and preference-aligned (RPO) Koel-TTS architectures, with and without CFG, on zero-shot TTS. \textbf{Right:} Equivalent evaluations for a \textit{Decoder Context} multilingual TTS model across various languages and text tokenizers. Both CFG and preference alignment, independently and in combination (green), improve CER and SSIM over the base model (gray).
    }
\vspace{-4mm}
    \label{figs:bar_graphs}
\end{figure*}

While CFG adds an inference-time cost by doubling the effective batch-size, it does not require any additional finetuning of the baseline models to achieve the substantial improvements across all metrics. 
Additionally, CFG inference on a preference aligned model results in further improvements and yields the best results across all metrics.
Figure~\ref{figs:bar_graphs} presents these improvements on unseen speaker TTS task across all models. Table~\ref{tab:dpo_rpo_cfg_base_multi} shows the improvements obtained with CFG on both seen and unseen speakers for the multi encoder architecture, on the baseline and preference aligned model. 
The reduction in CER/WER confidence intervals indicates that we can generate accurate speech reliably.

\subsection{Multilingual TTS}
\label{sec:multilingualexperiment}
For multilingual TTS, we investigate six languages English, Spanish, German, French, Italian, and Dutch.
For non-English languages, we use the CML dataset~\cite{oliveira2023cml} that contains $1{,}562$ hours of German, $642$ hours of Dutch, $476$ hours of Spanish, $283$ hours of French, $131$ hours of Italian speech data. Additionally, we incorporate $42$ hours of internal Spanish data from two speakers. Combining this with our $18k$ hours of English TTS data, we create a final blend of $21k$ hours of multilingual TTS data. We use the decoder context architecture for this task, and scale-up our model to a $1.1$ billion parameter model (See Appendix~\ref{sec:trainingdetails} for details).

For preference alignment of the multilingual model, we create $10k$ text and context audio pairs per language (by pairing texts with a random context audio), from the CML training data of each language. We combine these pairs with $20k$ English text and context audio pairs randomly sampled from the $58k$ pairs used in our primary experiments. We utilize the \textit{whisper-large-v3}~\cite{radford2022whisper} ASR model in our reward system to create preference pairs and perform DPO finetuning with $\beta$\texttt{=0.05}.

Figure~\ref{figs:bar_graphs}(right) presents the results of multilingual-TTS evaluations on unseen speakers from each language. For evaluations on non-English languages, we use $100$
utterances per language from the CML test set of each language. For English evaluation, we use the same $180$ utterances from LibriTTS \textit{test-clean} subset as used in our primary experiments. 
As shown by the results, both preference alignment and CFG (with $\gamma$\texttt{=2.5}) yield substantial improvement in both intelligibility and speaker similarity metrics across various languages and tokenizers.
More interestingly, CFG inference on a DPO finetuned checkpoint, yields substantial speaker similarity improvements over using either DPO or CFG in isolation, especially for non-English languages.

We find that Koel-TTS can work effectively on raw character tokens, and achieve similar results as using phonetic inputs, for languages in which we consider both phoneme and character tokenizers (English, Spanish and German). 
Incorporating both CFG and DPO, our multilingual model achieves similar CER as the English-only decoder context model and slightly improves speaker similarity (0.740 vs. 0.726).
We present ablations with alternate multilingual tokenization schemes in Appendix~\ref{sec:multilingualablations}.


\setlength{\tabcolsep}{2pt}
\begin{table}[ht]
\centering
\vspace{-4mm}
\caption{\footnotesize{Intelligibility, SSIM and naturalness evaluation of various zero-shot TTS models on a subset of \textit{test-clean} LibriTTS data.}}
\resizebox{1.0\columnwidth}{!}{%
\begin{tabular}{l|ccc|cc}
\toprule
Model & CER (\%) & $\downarrow$ WER (\%) $\downarrow$ & SSIM $\uparrow$ & MOS $\uparrow$ & SMOS $\uparrow$\\
\midrule
Ground Truth & $0.80$ & $1.83$ & $0.771$ & $3.937 \pm 0.028$ & - \\
VALLE-X~\cite{zhang2023speak} & $6.65$ & $11.28$ & $0.679$ & $3.532 \pm 0.046$ & $3.709 \pm 0.045$ \\
YourTTS~\cite{casanova2022yourtts} & $2.44$ & $5.19$ & $0.581$ & $3.235 \pm 0.047$ & $3.229 \pm 0.053$ \\
T5-TTS~\cite{t5tts} & $1.66$ & $3.28$ & $0.459$ & $3.533 \pm 0.046$ & $3.366 \pm 0.050$\\
E2-TTS~\cite{eskimez2024e2} & $1.29$ & $2.66$ & $\mathbf{0.848}$ & $3.889 \pm 0.040$ & $3.793 \pm 0.045$\\
F5-TTS~\cite{chen2024f5} & $1.23$ & $2.55$ & $0.834$ & $3.930 \pm 0.042$ & $3.785 \pm 0.045$\\
XTTS-v2~\cite{casanova2024xtts} & $0.99$ & $2.09$ & $0.680$ & $3.715 \pm 0.043$ & $3.434 \pm 0.050$\\
StyleTTS-2~\cite{li2024styletts} & $0.75$ & $1.52$ & $0.579$ & $4.047 \pm 0.039$ & $3.786 \pm 0.044$ \\
\midrule
Koel-TTS 380m English & $\mathbf{0.55}$ & $\mathbf{1.41}$ & $0.726$ & $4.054 \pm 0.039$ & $3.826 \pm 0.044$\\
Koel-TTS 1.1b Multilingual & $0.63$ & $1.42$ & $0.740$ &  $\mathbf{4.058 \pm 0.040}$ & $\mathbf{3.848 \pm 0.043}$ \\
\bottomrule
\end{tabular}
}
\vspace{-4mm}
\label{tab:results_libri_unseen_test}
\end{table}

\begin{figure}[!ht]
    \centering
    \includegraphics[width=1.0\columnwidth]{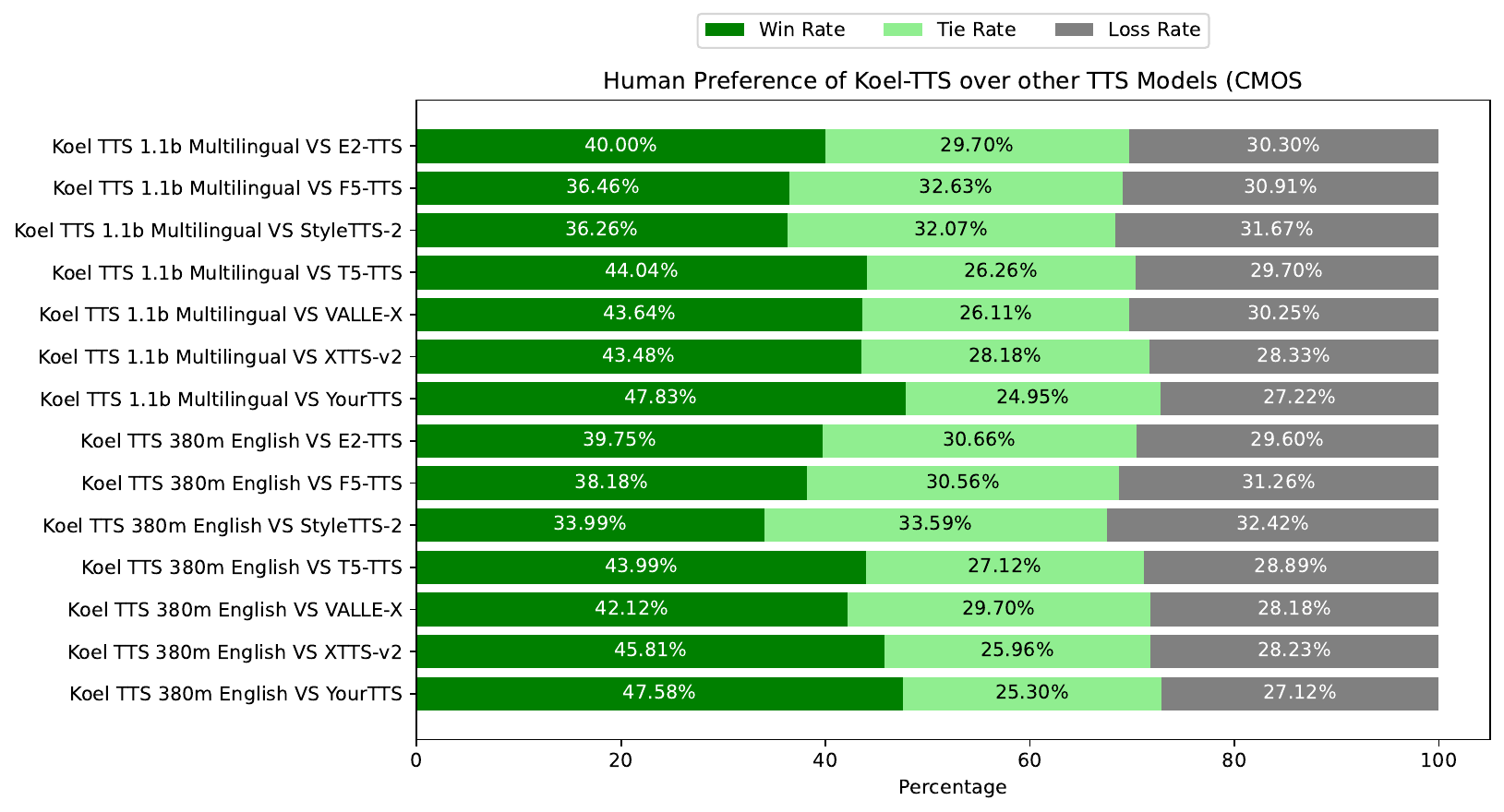}
    \vspace{-8mm} 
    \caption{\footnotesize{Koel-TTS vs. Previous Models: Dark green bars indicate the percentage of instances where human listeners preferred Koel-TTS for audio naturalness during side-by-side evaluations.}}
    \label{figs:cmosplot}
     \vspace{-4mm} 
\end{figure}

\subsection{Comparison against Past Work}
\label{sec:pastworkcomparison}
We benchmark two candidate Koel-TTS models: English-only decoder context model and the larger multilingual decoder context model against past work and open source models. We evaluate all models for zero-shot TTS on the unseen speaker evaluation set (test-clean LibriTTS subset), using the same evaluation procedure as described in Section~\ref{sec:datasets}.
We also compute three human evaluation metrics on Amazon Mechanical Turk namely Naturalness Mean Opinion Score (MOS), Speaker similarity MOS (SMOS) and Comparative MOS (CMOS). For complete details on MOS studies, see Appendix~\ref{sec:humaneval}. 
As shown in Table~\ref{tab:results_libri_unseen_test}, Koel-TTS achieves state-of-the-art intelligibility scores (CER/WER) despite being trained on significantly less data than competing models. While Koel-TTS outperforms LLM-based baselines (VALLE-X and XTTS-v2) in SSIM scores, it slightly underperforms CFM-based systems (F5-TTS and E2-TTS), which leverage 100k+ hours of speech data, compared to $21k$ hours for our largest model. Human evaluations of naturalness (MOS) and speaker similarity (SMOS) 
show Koel-TTS to be equally or more preferred compared to all other models. We attribute the difference between SSIM scores and SMOS to SSIM’s emphasis on timbre similarity, whereas human ratings consider additional factors such as style and accent. CMOS results in Figure~\ref{figs:cmosplot}, further confirm that Koel-TTS is preferred over all competing approaches.

%% file: sections/conclusion.tex
We introduce Koel-TTS, a suite of encoder-decoder transformer models that map text and context audio to acoustic speech tokens. 
By incorporating preference alignment driven by transcription accuracy and speaker similarity, and Classifier Free Guidance,
we improve intelligibility, speaker similarity, and naturalness of generated speech,  achieving state-of-the-art zero-shot TTS performance. 
Koel-TTS excels in multilingual TTS, delivering high-quality, low-latency speech with a simplified model design. Finally, through audio examples on our webpage, we demonstrate that Koel-TTS can be effectively fine-tuned for long-form multi-turn dialogue generation, by adapting the model for contextually aware podcast synthesis.

%% file: appendix.tex
\section{Pareto optimal ranking for creating preference pairs}
\label{paretorankingalgo}
Pareto optimal ranking is a technique for multi-attribute decision making~\cite{deb2011multi}. The key idea is to find non-dominated solutions and removing them from the current set recursively till we have ranked all items. When we find multiple items in the same pareto front, we break the ties by prioritizing our preference for more robust examples (lower CER), and we break any remaining ties by preferring higher SSIM. Below is the python code for ranking for this procedure.

\begin{lstlisting}[style=mypython, caption=Pareto Optimal Ranking of generated outputs for a given text-context pair using CER and SSIM metrics]
def pareto_ranking(items):
    """
    Given a list of (cer, ssim, item_idx), return the list of items
    sorted by their Pareto rank (rank 1 is best). Items in the same
    rank are sorted by ascending cer and incase of a tie, by descending ssim.
    
    :param items: List of tuples (cer, ssim, item_idx).
    :return: A list of tuples (rank, cer, ssim, item_idx), sorted first by rank,
             then by ascending cer within the same rank.
    """
    
    # A helper function to check if item A is dominated by item B
    # A: (cerA, ssimA), B: (cerB, ssimB)
    def is_dominated(A, B):
        return (B[0] <= A[0]) and (B[1] >= A[1]) and (B != A)
    
    remaining = items[:]
    
    ranked_items = []  # Will hold tuples of (rank, cer, ssim, item_idx)
    current_rank = 1
    
    while remaining:
        # Find all non-dominated items in the current set 'remaining'
        non_dominated = []
        for i in range(len(remaining)):
            dominated = False
            for j in range(len(remaining)):
                if i != j:
                    if is_dominated(remaining[i], remaining[j]):
                        dominated = True
                        break
            if not dominated:
                non_dominated.append(remaining[i])
        
        # Assign current_rank to all non-dominated items
        # and remove them from remaining
        for nd in non_dominated:
            ranked_items.append((current_rank, nd[0], nd[1], nd[2]))
            remaining.remove(nd)
        
        current_rank += 1
    
    # Now sort the ranked items by (rank asc, cer asc, ssim desc)
    ranked_items.sort(key=lambda x: (x[0], x[1], -x[2]))
    
    return ranked_items
\end{lstlisting}

\setlength{\tabcolsep}{4pt}
\begin{table*}[t]
\vspace{-2mm}
\caption{\footnotesize{Evaluation of DPO, RPO and CFG on baseline models for all Koel-TTS architectures. We consider two DPO experiments with $\beta=(0.01,0.05)$. 
}}
\centering
\resizebox{\textwidth}{!}{%
\begin{tabular}{l|cccc|cccc}
\multicolumn{1}{c}{} & \multicolumn{4}{c}{\emph{Seen Speakers}} & \multicolumn{4}{c}{\emph{Unseen Speakers}} \\
\toprule
Model/Technique & CER(\%) $\downarrow$ & WER(\%) $\downarrow$ & SSIM $\uparrow$ & Squim-MOS $\uparrow$ & CER(\%) $\downarrow$ & WER(\%) $\downarrow$ & SSIM $\uparrow$ & Squim-MOS $\uparrow$ \\
\midrule
Multi Encoder (BL-1) & $1.92 \pm 0.68$ & $3.02 \pm 0.76$ & $0.712 \pm 0.002$ & $4.346 \pm 0.028$ & $2.56 \pm 1.44$ & $3.74 \pm 1.36$ & $0.601 \pm 0.004$ & $4.318 \pm 0.059$ \\ 
BL-1 + RPO ($\beta=0.01$) & $1.01 \pm 0.58$ & $1.76 \pm 0.59$ & $0.737 \pm 0.002$ & $4.408 \pm 0.010$ & $0.79 \pm 0.12$ & $1.72 \pm 0.18$ & $0.641 \pm 0.002$ & $4.389 \pm 0.021$ \\ 
BL-1 + DPO ($\beta=0.01$) & $0.67 \pm 0.17$ & $1.48 \pm 0.34$ & $0.737 \pm 0.004$ & $4.406 \pm 0.011$ & $0.62 \pm 0.12$ & $1.49 \pm 0.20$ & $0.645 \pm 0.001$ & $4.402 \pm 0.029$ \\ 
BL-1 + DPO ($\beta=0.05$) & $1.30 \pm 0.51$ & $2.25 \pm 0.53$ & $0.737 \pm 0.003$ & $4.401 \pm 0.014$ & $1.01 \pm 0.40$ & $2.01 \pm 0.49$ & $0.643 \pm 0.005$ & $4.402 \pm 0.013$ \\ 
BL-1 + CFG ($\gamma=2.5$) & $0.73 \pm 0.16$ & $1.63 \pm 0.24$ & $0.752 \pm 0.003$ & $4.420 \pm 0.010$ & $0.69 \pm 0.07$ & $1.59 \pm 0.10$ & $0.653 \pm 0.005$ & $4.415 \pm 0.007$ \\ 
BL-1 + RPO + CFG ($\gamma=2.5$)& $0.75 \pm 0.23$ & $1.56 \pm 0.32$ & $0.766 \pm 0.003$ & $\mathbf{4.422 \pm 0.011}$ & $0.51 \pm 0.12$ & $1.25 \pm 0.18$ & $0.674 \pm 0.004$ & $4.392 \pm 0.029$ \\ 
BL-1 + DPO ($\beta=0.01$) + CFG ($\gamma=2.5$)& $\mathbf{0.51 \pm 0.12}$ & $\mathbf{1.32 \pm 0.23}$ & $\mathbf{0.767 \pm 0.004}$ & $4.418 \pm 0.012$ & $0.58 \pm 0.17$ & $1.38 \pm 0.08$ & $\mathbf{0.678 \pm 0.005}$ & $\mathbf{4.417 \pm 0.015}$ \\ 
BL-1 + DPO ($\beta=0.05$) + CFG ($\gamma=2.5$)& $1.12 \pm 0.83$ & $1.87 \pm 0.85$ & $0.766 \pm 0.002$ & $4.420 \pm 0.011$ & $\mathbf{0.49 \pm 0.07}$ & $\mathbf{1.24 \pm 0.11}$ & $0.676 \pm 0.004$ & $4.390 \pm 0.025$ \\ 
\midrule
Decoder Context (BL-2) & $1.73 \pm 0.60$ & $2.98 \pm 0.59$ & $0.700 \pm 0.001$ & $4.350 \pm 0.038$ & $2.68 \pm 1.13$ & $4.02 \pm 1.12$ & $0.637 \pm 0.008$ & $4.347 \pm 0.024$ \\ 
BL-2 + RPO ($\beta=0.01$) & $1.01 \pm 0.60$ & $2.03 \pm 0.62$ & $0.719 \pm 0.002$ & $4.403 \pm 0.013$ & $1.17 \pm 0.94$ & $2.09 \pm 1.00$ & $0.681 \pm 0.005$ & $4.401 \pm 0.012$ \\ 
BL-2 + DPO ($\beta=0.01$) & $1.32 \pm 0.40$ & $2.39 \pm 0.46$ & $0.708 \pm 0.004$ & $4.392 \pm 0.017$ & $0.89 \pm 0.15$ & $1.90 \pm 0.28$ & $0.667 \pm 0.003$ & $4.400 \pm 0.012$ \\ 
BL-2 + DPO ($\beta=0.05$) & $1.25 \pm 0.83$ & $2.27 \pm 0.97$ & $0.716 \pm 0.004$ & $4.393 \pm 0.016$ & $0.98 \pm 0.46$ & $2.03 \pm 0.49$ & $0.676 \pm 0.004$ & $4.408 \pm 0.010$ \\ 
BL-2 + CFG ($\gamma=2.5$) & $0.62 \pm 0.20$ & $1.58 \pm 0.44$ & $0.741 \pm 0.003$ & $\mathbf{4.418 \pm 0.009}$ & $0.57 \pm 0.11$ & $\mathbf{1.37 \pm 0.11}$ & $0.720 \pm 0.004$ & $\mathbf{4.417 \pm 0.007}$ \\ 
BL-2 + RPO + CFG ($\gamma=2.5$)& $\mathbf{0.51 \pm 0.12}$ & $\mathbf{1.38 \pm 0.25}$ & $\mathbf{0.751 \pm 0.002}$ & $4.415 \pm 0.013$ & $\mathbf{0.55 \pm 0.11}$ & $1.41 \pm 0.19$ & $\mathbf{0.729 \pm 0.003}$ & $4.415 \pm 0.012$ \\ 
BL-2 + DPO ($\beta=0.01$) + CFG ($\gamma=2.5$)& $0.62 \pm 0.09$ & $1.53 \pm 0.17$ & $0.744 \pm 0.002$ & $4.409 \pm 0.019$ & $0.60 \pm 0.10$ & $1.40 \pm 0.31$ & $0.720 \pm 0.001$ & $4.387 \pm 0.038$ \\ 
BL-2 + DPO ($\beta=0.05$) + CFG ($\gamma=2.5$)& $0.54 \pm 0.08$ & $1.43 \pm 0.19$ & $0.749 \pm 0.005$ & $4.413 \pm 0.018$ & $0.55 \pm 0.10$ & $1.42 \pm 0.28$ & $\mathbf{0.729 \pm 0.003}$ & $4.413 \pm 0.013$ \\ 
\midrule
SV Conditioned (BL-3) & $1.71 \pm 0.41$ & $2.82 \pm 0.41$ & $0.697 \pm 0.003$ & $4.360 \pm 0.021$ & $3.12 \pm 0.98$ & $4.22 \pm 1.02$ & $0.619 \pm 0.003$ & $4.318 \pm 0.034$ \\ 
BL-3 + RPO ($\beta=0.01$) & $0.72 \pm 0.14$ & $1.61 \pm 0.22$ & $0.717 \pm 0.001$ & $4.408 \pm 0.010$ & $1.25 \pm 0.38$ & $2.06 \pm 0.49$ & $0.668 \pm 0.003$ & $4.389 \pm 0.026$ \\ 
BL-3 + DPO ($\beta=0.01$) & $0.62 \pm 0.19$ & $1.46 \pm 0.31$ & $0.705 \pm 0.003$ & $4.402 \pm 0.011$ & $0.76 \pm 0.12$ & $1.67 \pm 0.11$ & $0.650 \pm 0.003$ & $4.384 \pm 0.023$ \\ 
BL-3 + DPO ($\beta=0.05$) & $1.24 \pm 0.33$ & $2.19 \pm 0.33$ & $0.713 \pm 0.002$ & $4.416 \pm 0.042$ & $1.64 \pm 0.70$ & $2.64 \pm 0.69$ & $0.663 \pm 0.003$ & $4.385 \pm 0.019$ \\ 
BL-3 + CFG ($\gamma=2.5$) & $0.48 \pm 0.12$ & $1.38 \pm 0.33$ & $0.738 \pm 0.003$ & $4.407 \pm 0.025$ & $0.52 \pm 0.11$ & $1.38 \pm 0.20$ & $0.703 \pm 0.002$ & $\mathbf{4.418 \pm 0.011}$ \\ 
BL-3 + RPO + CFG ($\gamma=2.5$)& $0.46 \pm 0.10$ & $1.25 \pm 0.14$ & $\mathbf{0.750 \pm 0.003}$ & $\mathbf{4.423 \pm 0.010}$ & $0.57 \pm 0.15$ & $1.33 \pm 0.21$ & $\mathbf{0.715 \pm 0.003}$ & $4.416 \pm 0.014$ \\ 
BL-3 + DPO ($\beta=0.01$) + CFG ($\gamma=2.5$)& $0.46 \pm 0.05$ & $\mathbf{1.24 \pm 0.11}$ & $0.743 \pm 0.002$ & $4.417 \pm 0.015$ & $\mathbf{0.47 \pm 0.07}$ & $\mathbf{1.26 \pm 0.09}$ & $0.706 \pm 0.004$ & $4.412 \pm 0.014$ \\ 
BL-3 + DPO ($\beta=0.05$) + CFG ($\gamma=2.5$)& $\mathbf{0.45 \pm 0.13}$ & $1.27 \pm 0.06$ & $0.747 \pm 0.001$ & $4.403 \pm 0.021$ & $0.70 \pm 0.45$ & $1.51 \pm 0.44$ & $\mathbf{0.715 \pm 0.004}$ & $4.373 \pm 0.055$ \\ 
\bottomrule
\end{tabular} 
}
\label{tab:dporpoallmodels}
\end{table*}
\vspace{-5mm}

\section{DPO and RPO on all model architectures}
\label{sec:dporpoall}
We perform DPO and RPO preference optimization on all models and evaluate the preference aligned checkpoint with and without CFG. Results are reported in Table~\ref{tab:dporpoallmodels}. We observe a significant improvement in CER, WER and SSIM across all baseline models, when preference alignment or CFG is done in isolation. Across most architectures, the best metrics are achieved by CFG inference on a preference aligned checkpoint (DPO + CFG or RPO + CFG). 
Both DPO and RPO perform similarly, but in practice, we find DPO to be more sensitive to $\beta$ hyperparameter as compared to RPO.

\section{MOS, SMOS and CMOS Evaluation}
\label{sec:humaneval}
\begin{figure}[h]
    \centering
    \begin{minipage}{0.31\textwidth}
        \centering
        \includegraphics[width=\linewidth]{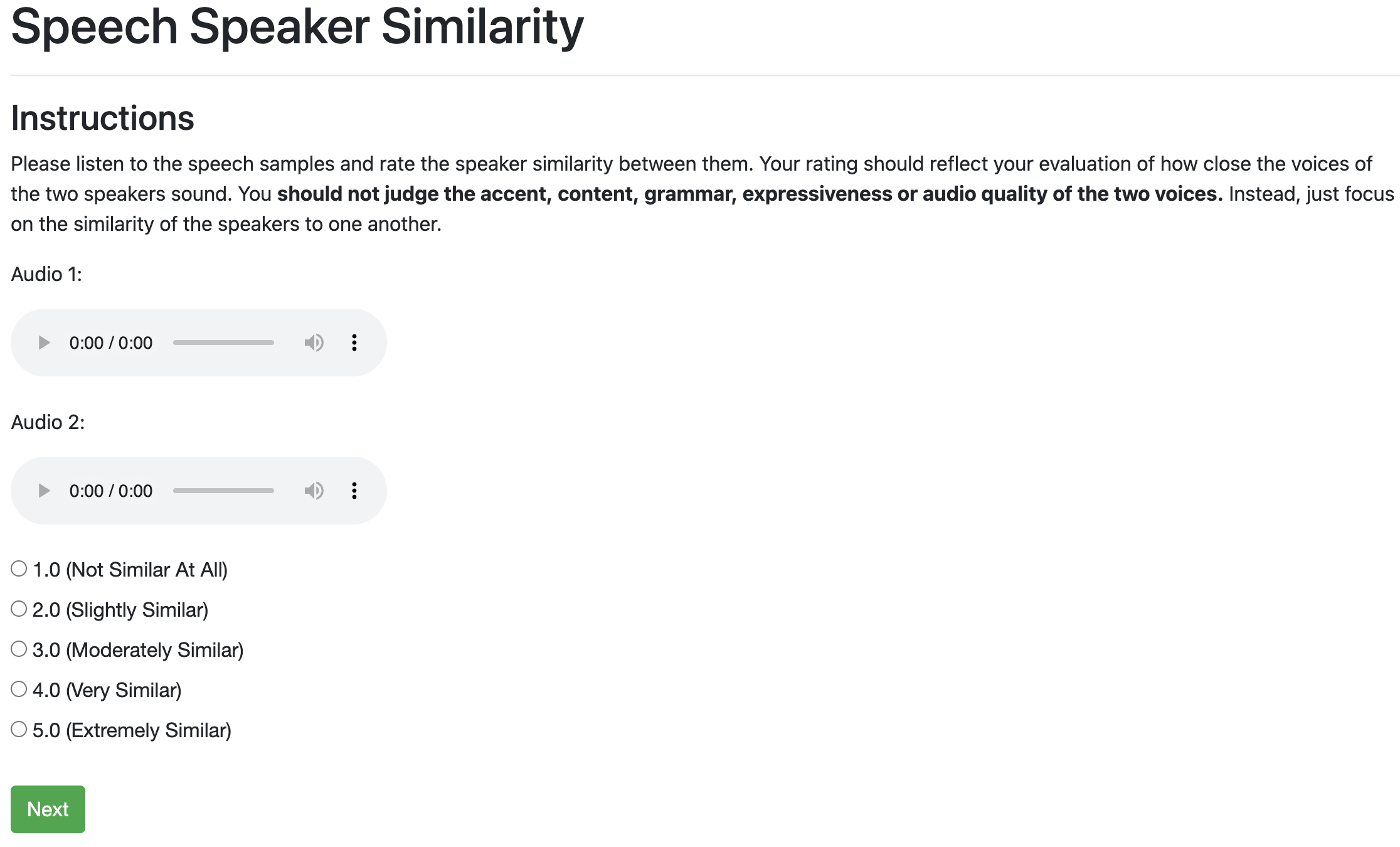}
        \caption{\footnotesize{User Study template used for Speaker Similarity (SMOS) evaluation}}
        \label{figs:simmostemplate}
    \end{minipage}\hfill
    \begin{minipage}{0.31\textwidth}
        \centering
        \includegraphics[width=\linewidth]{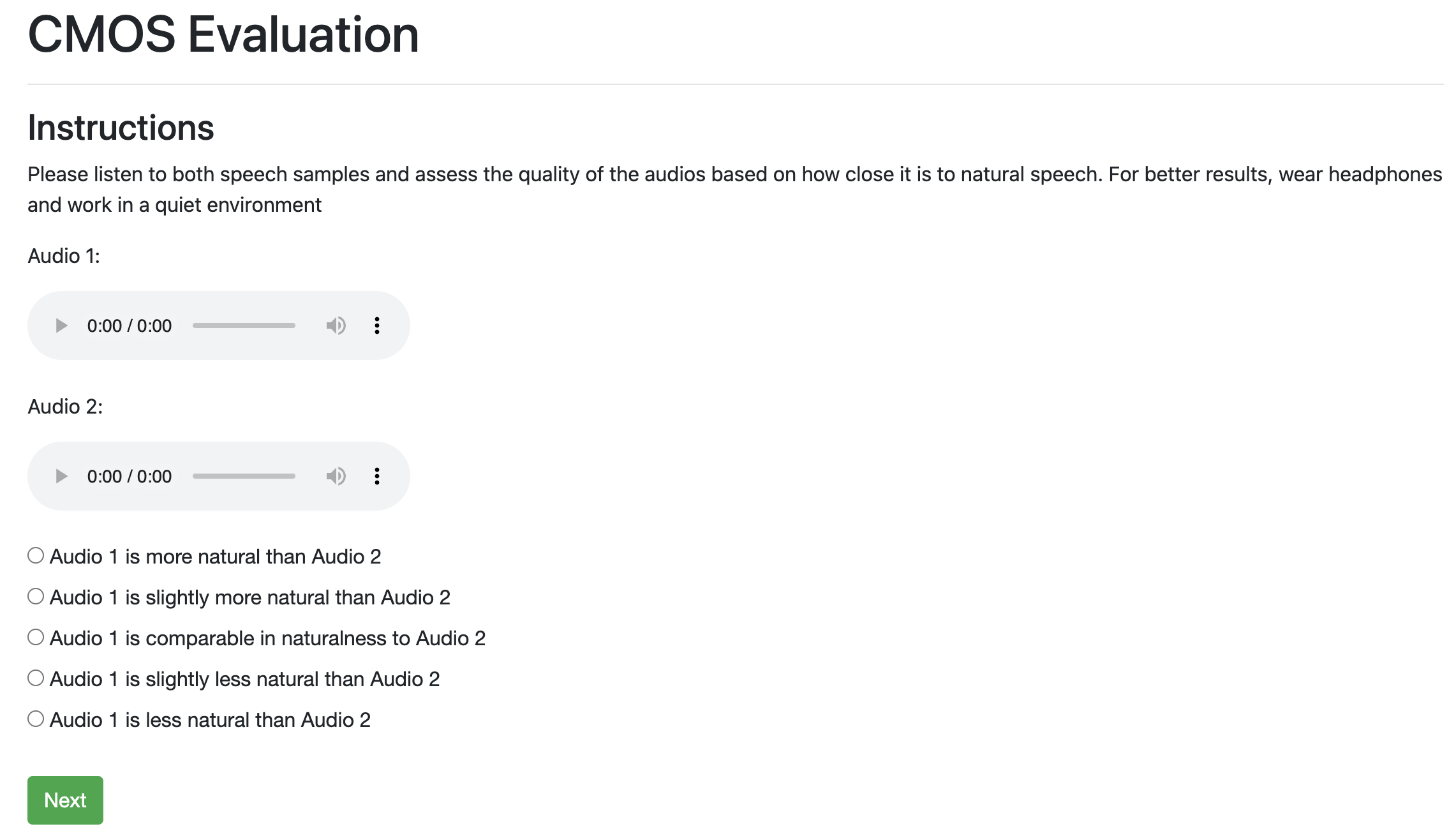}
        \vspace{2mm}
        \caption{\footnotesize{User Study template used for Comparative-CMOS evaluation}}
        \label{figs:cmostemplate}
    \end{minipage}\hfill
    \begin{minipage}{0.31\textwidth}
        \centering
        \includegraphics[width=\linewidth]{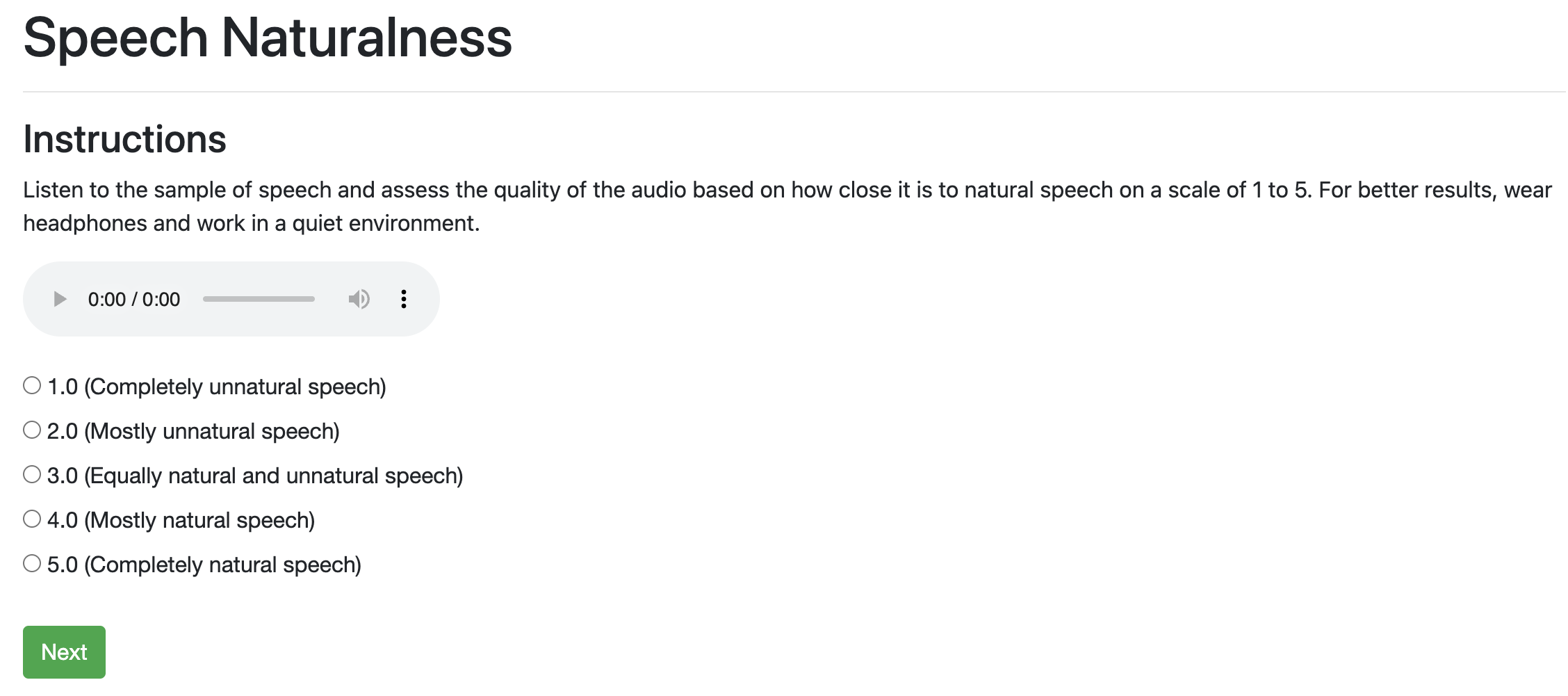}
        \vspace{6mm}
        \caption{\footnotesize{User Study template used for MOS evaluation}}
        \label{figs:mostemplate}
    \end{minipage}
\end{figure}

\textbf{Naturalness MOS Evaluation:} We ask human listeners to rate the audio on a scale of $1$ to $5$ point naturalness scale with $1$ point increments. We present $180$ audio examples of each technique and each audio is independently rated by at least $11$ listeners. This results in a total of $1980$ evaluations per technique. 
The template used for the Naturalness human study is shown in Figure~\ref{figs:mostemplate}. We report the MOS with $95\%$ confidence intervals in Table~\ref{tab:results_libri_unseen_test} of the paper.

\textbf{Speaker Similarity MOS (SMOS):} For SMOS evaluation, we ask human listeners to rate the speaker similarity of a given pair of utterances. For this evaluation, each synthetic utterance is paired with a real context utterance of the target speaker. We create pairs for all of the $180$ synthesized utterances of each technique. Each pair is rated by at least $11$ independent listeners resulting in at least $1800$ speaker similarity evaluations of each technique. We ask the listeners to judge only the voice/speaker of the utterances and ignore the accent, content, grammar and expressiveness of speech. following past work~\citep{casanova2022yourtts,hussain2023ace,neekharaselfvc}. The templates used for this user study are shown in Figures~\ref{figs:simmostemplate}, \ref{figs:cmostemplate} and \ref{figs:mostemplate}. 

\textbf{Comparative MOS (CMOS):} For CMOS, listeners are asked to compare two audio utterances on naturalness and indicate their preference as one of the five options shown in Figure~\ref{figs:cmostemplate}. We pair the two Koel-TTS models with all other models. We evaluate the percentage of times across $1800$ evaluations that Koel-TTS is preferred over an alternate model.

\section{Multilingual Tokenization Ablations}
\label{sec:multilingualablations}
We train three decoder-context Koel-TTS models considering three tokenization schemes besides phonemes --- Model A: Aggregated characters from different languages (Vocab size $=256 \times \text{Number of Languages}$). Model A is the default model in our primary multilingual experiments. Model B: Shared character tokenizer (256 character tokens shared across all languages). Model C: Multilingual sentence piece tokens~\footnote{ \url{https://huggingface.co/google-bert/bert-base-multilingual-uncased}} (Vocab size 110k). We find that character-based tokenizers perform significantly better than sentence piece tokenizer on intelligibility metrics, especially when unseen words are encountered during inference. Table~\ref{tab:multilingualablations} compares the different models for each language studied in our work. 
All results are reported using CFG scale $\gamma=2.5$, without any preference alignment. 
Additionally, we find the aggregated char tokenizer to perform better for cross-lingual TTS synthesis (when the context audio has a different language than the input text). This is because token embeddings for each language are independent from the others and not shared (as in the case of the shared character tokenizer).

\setlength{\tabcolsep}{6pt}
\begin{table}
\vspace{-2mm}
\caption{\footnotesize{Comparison of decoder context Koel-TTS models trained using different text tokenizers, considering all allowed tokenizations at test time. (CFG Scale $\gamma=2.5$, No Preference Alignment). Evaluation conduced on unseen speakers for each language on the test set described in Section~\ref{sec:multilingualexperiment}}}
\centering
\resizebox{1.0\textwidth}{!}{%
\begin{tabular}{lll|cccc}
\toprule
Model & Language & Tokenizer & CER(\%) $\downarrow$ & WER(\%) $\downarrow$ & SSIM $\uparrow$  \\
\midrule
Model A (Phoneme + Aggregated characters) & English & Phonemes & $0.78 \pm 0.77$ & $1.65 \pm 0.85$ & $0.735 \pm 0.002$ \\
Model B (Phoneme + Multilingual sentencepiece) & English & Phonemes & $0.59 \pm 0.10$ & $1.57 \pm 0.21$ & $0.746 \pm 0.003$ \\
Model C (Phoneme + Shared characters) & English & Phonemes & $0.60 \pm 0.20$ & $1.41 \pm 0.23$ & $0.739 \pm 0.002$ \\
Model B (Phoneme + Multilingual sentencepiece) & English & Multilingual sentencepiece & $0.58 \pm 0.09$ & $1.44 \pm 0.17$ & $\mathbf{0.747 \pm 0.000}$ \\
Model C (Phoneme + Shared characters) & English & Shared characters & $\mathbf{0.52 \pm 0.04}$ & $\mathbf{1.37 \pm 0.08}$ & $0.739 \pm 0.001$ \\
Model A (Phoneme + Aggregated characters) & English & Aggregated characters & $0.77 \pm 0.49$ & $1.67 \pm 0.47$ & $0.731 \pm 0.005$ \\
\midrule
Model A (Phoneme + Aggregated characters) & Spanish & Phonemes & $1.37 \pm 1.53$ & $3.63 \pm 1.99$ & $0.693 \pm 0.013$ \\
Model B (Phoneme + Multilingual sentencepiece) & Spanish & Phonemes & $1.52 \pm 0.71$ & $4.01 \pm 0.35$ & $0.698 \pm 0.011$ \\
Model C (Phoneme + Shared characters) & Spanish & Phonemes & $1.41 \pm 0.56$ & $3.73 \pm 0.59$ & $0.703 \pm 0.007$ \\
Model B (Phoneme + Multilingual sentencepiece) & Spanish & Multilingual sentencepiece & $1.96 \pm 0.68$ & $4.93 \pm 0.67$ & $0.701 \pm 0.007$ \\
Model C (Phoneme + Shared characters) & Spanish & Shared characters & $1.64 \pm 0.81$ & $4.03 \pm 1.08$ & $\mathbf{0.704 \pm 0.011}$ \\
Model A (Phoneme + Aggregated characters) & Spanish & Aggregated characters & $\mathbf{1.03 \pm 0.11}$ & $\mathbf{3.11 \pm 0.18}$ & $0.693 \pm 0.008$ \\
\midrule
Model A (Phoneme + Aggregated characters) & German & Phonemes & $2.69 \pm 2.36$ & $5.75 \pm 3.25$ & $0.607 \pm 0.008$ \\
Model B (Phoneme + Multilingual sentencepiece) & German & Phonemes & $4.01 \pm 3.89$ & $6.88 \pm 5.05$ & $0.613 \pm 0.021$ \\
Model C (Phoneme + Shared characters) & German & Phonemes & $\mathbf{1.67 \pm 0.71}$ & $\mathbf{4.54 \pm 1.05}$ & $\mathbf{0.645 \pm 0.007}$ \\
Model B (Phoneme + Multilingual sentencepiece) & German & Multilingual sentencepiece & $4.28 \pm 3.13$ & $7.77 \pm 2.90$ & $0.611 \pm 0.014$ \\
Model C (Phoneme + Shared characters) & German & Shared characters & $1.93 \pm 1.06$ & $4.76 \pm 1.08$ & $0.644 \pm 0.005$ \\
Model A (Phoneme + Aggregated characters) & German & Aggregated characters & $2.02 \pm 1.13$ & $4.81 \pm 0.95$ & $0.614 \pm 0.003$ \\
\midrule
Model B (Phoneme + Multilingual sentencepiece) & French & Multilingual sentencepiece & $6.77 \pm 2.12$ & $10.33 \pm 2.17$ & $0.638 \pm 0.009$ \\
Model C (Phoneme + Shared characters) & French & Shared characters & $2.30 \pm 0.70$ & $\mathbf{5.35 \pm 0.81}$ & $\mathbf{0.643 \pm 0.006}$ \\
Model A (Phoneme + Aggregated characters) & French & Aggregated characters & $\mathbf{2.28 \pm 1.26}$ & $5.50 \pm 1.57$ & $0.641 \pm 0.011$ \\
\midrule
Model B (Phoneme + Multilingual sentencepiece) & Italian & Multilingual sentencepiece & $3.68 \pm 0.65$ & $12.83 \pm 1.07$ & $\mathbf{0.650 \pm 0.003}$ \\
Model C (Phoneme + Shared characters) & Italian & Shared characters & $4.99 \pm 1.83$ & $12.81 \pm 2.15$ & $0.649 \pm 0.009$ \\
Model A (Phoneme + Aggregated characters) & Italian & Aggregated characters & $\mathbf{2.37 \pm 0.50}$ & $\mathbf{9.64 \pm 1.09}$ & $0.615 \pm 0.003$ \\
\midrule
Model B (Phoneme + Multilingual sentencepiece) & Dutch & Multilingual sentencepiece & $4.19 \pm 1.24$ & $11.49 \pm 1.96$ & $0.607 \pm 0.004$ \\
Model C (Phoneme + Shared characters) & Dutch & Shared characters & $\mathbf{3.05 \pm 0.93}$ & $\mathbf{9.79 \pm 1.57}$ & $\mathbf{0.613 \pm 0.008}$ \\
Model A (Phoneme + Aggregated characters) & Dutch & Aggregated characters & $3.32 \pm 1.51$ & $10.14 \pm 1.18$ & $0.594 \pm 0.006$ \\
\bottomrule
\end{tabular} 
}
\label{tab:multilingualablations}
\end{table}

\section{Model Architecture and Training Hyperparameters}
\label{sec:trainingdetails}
For our primary experiments (Table~\ref{tab:architecture_comp}, Table~\ref{tab:dpo_rpo_cfg_base_multi} and Table~\ref{tab:dporpoallmodels}), in each architecture, the decoder is built with $12$ transformer layers using a hidden dimension of $768$ and a feed-forward network (FFN) dimension of $3072$. Rather than a standard FFN sublayer, this decoder employs a causal convolution layer with a kernel size of $3$. The transcript encoder, on the other hand, is composed of $6$ transformer layers that do not use causal masking but otherwise match the decoder's specifications. For multi encoder, a $3$ layer context encoder is added, which uses the same non-causal design as the transcript encoder. In the multi encoder architecture, the transcript encoding is fed to cross attention layers of decoder layer ($[0,2,4,6,8,10]$) and context audio tokens are fed to decoder layers ($[1,3,5,7,9,11]$). For all other models, transcript encoding goes to all decoder layers. Our models are trained on $16$ NVIDIA A100 GPUs using a global batch size of $256$, optimized using Adam optimizer with an initial learning rate of $1e-4$. The learning rate is annealed every $1000$ training steps using an exponential decay factor of $0.998$. 
Training for different architectures on English converges in around $200k$ steps with this configuration and takes around $40$ hours. 

For our larger multilingual model, we increase the hidden dimension to $1536$ and FFN dimension to $6144$ for both the encoder and decoder. The number of layers of the decoder are increased from $12$ to $16$, while keeping the same number of layers in the encoder. 
With this larger transformer, we set kernel size=$1$ for the decoder, and kernel size=$3$ for encoder. Training for this larger model converges in around $150k$ steps using a global batch size of $256$ across distributed across $32$ NVIDIA A100 GPUs.




\section{Evaluation on hard sentences with repeated words}
Autoregressive TTS models often struggle with challenging sentences containing repeated words. Issues such as infinite silences, looping of words become more prominent when presented with such challenging inputs. While cross-attention biasing~\cite{onealign,t5tts} partially addresses this issue, we find CFG and preference alignment can further improve robustness of the base model by mitigating hallucinations. 
Table~\ref{tab:challengingtexts} presents evaluation of base model and preference aligned + CFG models on a set of $91$ hard sentences linked in our webpage. We conduct these evaluations by pairing the challenging texts with $2$ seen speakers in our training dataset. As indicated by the results, while CFG and preference alignment improve CER and WER, there is scope for further improvement using inference-time monotonic alignment strategies.

\renewcommand{\arraystretch}{1.0} 
\setlength{\tabcolsep}{8pt}
\begin{table}[ht]

\caption{\footnotesize{Intelligibility and Speaker similarity evaluation on challenging sentences with repeated words on base models and the preference aligned models with CFG.}}

\centering
\resizebox{0.8\columnwidth}{!}{%
\begin{tabular}{l|cccccc}
\toprule
Model & CER(\%) $\downarrow$ & WER(\%) $\downarrow$ & SSIM $\uparrow$ \\
\midrule
Multi Encoder (Baseline) & $5.62 \pm 0.52$ & $10.60 \pm 0.90$ & $0.788 \pm 0.002$ \\
Multi Encoder (w Pref Align and CFG) & $\mathbf{5.03 \pm 0.16}$ & $\mathbf{9.19 \pm 0.46}$ & $\mathbf{0.807 \pm 0.001}$\\
\midrule
Decoder Context (Baseline) & $5.70 \pm 0.56$ & $10.38 \pm 0.35$ & $0.790 \pm 0.003$\\
Decoder Context (w Pref Align and CFG) & $\mathbf{4.69 \pm 0.14}$ &  $\mathbf{9.04 \pm 0.24}$ &  $\mathbf{0.798 \pm 0.002}$ \\
\midrule
SV Conditioned (Baseline) & $5.59 \pm 0.84$ & $10.33 \pm 0.94$ & $0.785 \pm 0.001$\\
SV Conditioned (w Pref Align and CFG) & $\mathbf{4.67 \pm 0.37}$ & $\mathbf{8.50 \pm 0.66}$ & $\mathbf{0.798 \pm 0.002}$ \\

\bottomrule
\end{tabular} 
}
\label{tab:challengingtexts}
\vspace{-2mm}
\end{table}


%% file: example_paper.bbl
\begin{thebibliography}{48}
\providecommand{\natexlab}[1]{#1}
\providecommand{\url}[1]{\texttt{#1}}
\expandafter\ifx\csname urlstyle\endcsname\relax
  \providecommand{\doi}[1]{doi: #1}\else
  \providecommand{\doi}{doi: \begingroup \urlstyle{rm}\Url}\fi

\bibitem[Adler et~al.(2024)Adler, Agarwal, Aithal, Anh, Bhattacharya, Brundyn, Casper, Catanzaro, Clay, Cohen, et~al.]{adler2024nemotron}
Adler, B., Agarwal, N., Aithal, A., Anh, D.~H., Bhattacharya, P., Brundyn, A., Casper, J., Catanzaro, B., Clay, S., Cohen, J., et~al.
\newblock Nemotron-4 340b technical report.
\newblock \emph{arXiv preprint arXiv:2406.11704}, 2024.

\bibitem[Azar et~al.(2024)Azar, Guo, Piot, Munos, Rowland, Valko, and Calandriello]{azar2024ipo}
Azar, M.~G., Guo, Z.~D., Piot, B., Munos, R., Rowland, M., Valko, M., and Calandriello, D.
\newblock A general theoretical paradigm to understand learning from human preferences.
\newblock In \emph{International Conference on Artificial Intelligence and Statistics}. PMLR, 2024.

\bibitem[Badlani et~al.(2022)Badlani, Łańcucki, Shih, Valle, Ping, and Catanzaro]{onealign}
Badlani, R., Łańcucki, A., Shih, K.~J., Valle, R., Ping, W., and Catanzaro, B.
\newblock One tts alignment to rule them all.
\newblock In \emph{ICASSP}, 2022.

\bibitem[Bakhturina et~al.(2021)Bakhturina, Lavrukhin, Ginsburg, and Zhang]{bakhturina21_interspeech}
Bakhturina, E., Lavrukhin, V., Ginsburg, B., and Zhang, Y.
\newblock {Hi-Fi Multi-Speaker English TTS Dataset}.
\newblock In \emph{INTERSPEECH}, 2021.

\bibitem[Borsos et~al.(2023)Borsos, Marinier, Vincent, Kharitonov, Pietquin, Sharifi, Roblek, Teboul, Grangier, Tagliasacchi, et~al.]{borsos2023audiolm}
Borsos, Z., Marinier, R., Vincent, D., Kharitonov, E., Pietquin, O., Sharifi, M., Roblek, D., Teboul, O., Grangier, D., Tagliasacchi, M., et~al.
\newblock Audiolm: a language modeling approach to audio generation.
\newblock \emph{IEEE/ACM Transactions on Audio, Speech, and Language Processing}, 2023.

\bibitem[Casanova et~al.(2022)Casanova, Weber, Shulby, Junior, G{\"o}lge, and Ponti]{casanova2022yourtts}
Casanova, E., Weber, J., Shulby, C.~D., Junior, A.~C., G{\"o}lge, E., and Ponti, M.~A.
\newblock Yourtts: Towards zero-shot multi-speaker tts and zero-shot voice conversion for everyone.
\newblock In \emph{International Conference on Machine Learning}. PMLR, 2022.

\bibitem[Casanova et~al.(2024)Casanova, Davis, G{\"o}lge, G{\"o}knar, Gulea, Hart, Aljafari, Meyer, Morais, Olayemi, et~al.]{casanova2024xtts}
Casanova, E., Davis, K., G{\"o}lge, E., G{\"o}knar, G., Gulea, I., Hart, L., Aljafari, A., Meyer, J., Morais, R., Olayemi, S., et~al.
\newblock Xtts: a massively multilingual zero-shot text-to-speech model.
\newblock \emph{INTERSPEECH}, 2024.

\bibitem[Casanova et~al.(2025)Casanova, Langman, Neekhara, Hussain, Li, Ghosh, Juki{\'c}, and Lee]{casanova2024low}
Casanova, E., Langman, R., Neekhara, P., Hussain, S., Li, J., Ghosh, S., Juki{\'c}, A., and Lee, S.-g.
\newblock Low frame-rate speech codec: a codec designed for fast high-quality speech llm training and inference.
\newblock \emph{ICASSP}, 2025.

\bibitem[Chen et~al.(2024{\natexlab{a}})Chen, Niu, Ma, Deng, Wang, Zhao, Yu, and Chen]{chen2024f5}
Chen, Y., Niu, Z., Ma, Z., Deng, K., Wang, C., Zhao, J., Yu, K., and Chen, X.
\newblock F5-tts: A fairytaler that fakes fluent and faithful speech with flow matching.
\newblock \emph{arXiv preprint arXiv:2410.06885}, 2024{\natexlab{a}}.

\bibitem[Chen et~al.(2024{\natexlab{b}})Chen, Deng, Yuan, Ji, and Gu]{chen2024self}
Chen, Z., Deng, Y., Yuan, H., Ji, K., and Gu, Q.
\newblock Self-play fine-tuning converts weak language models to strong language models.
\newblock \emph{arXiv preprint arXiv:2401.01335}, 2024{\natexlab{b}}.

\bibitem[Christiano et~al.(2017)Christiano, Leike, Brown, Martic, Legg, and Amodei]{christiano2017deep}
Christiano, P.~F., Leike, J., Brown, T., Martic, M., Legg, S., and Amodei, D.
\newblock Deep reinforcement learning from human preferences.
\newblock \emph{Advances in Neural Information Processing Systems}, 2017.

\bibitem[Darefsky et~al.(2024)Darefsky, Zhu, and Duan]{darefsky2024parakeet}
Darefsky, J., Zhu, G., and Duan, Z.
\newblock Parakeet, 2024.
\newblock URL \url{https://jordandarefsky.com/blog/2024/parakeet/}.

\bibitem[Deb(2011)]{deb2011multi}
Deb, K.
\newblock Multi-objective optimisation using evolutionary algorithms: an introduction.
\newblock In \emph{Multi-objective evolutionary optimisation for product design and manufacturing}. Springer, 2011.

\bibitem[D{\'e}fossez et~al.(2023)D{\'e}fossez, Copet, Synnaeve, and Adi]{encodec}
D{\'e}fossez, A., Copet, J., Synnaeve, G., and Adi, Y.
\newblock High fidelity neural audio compression.
\newblock \emph{Transactions on Machine Learning Research}, 2023.

\bibitem[Du et~al.(2024)Du, Wang, Chen, Shi, Lv, Zhao, Gao, Yang, Gao, Wang, et~al.]{du2024cosyvoice}
Du, Z., Wang, Y., Chen, Q., Shi, X., Lv, X., Zhao, T., Gao, Z., Yang, Y., Gao, C., Wang, H., et~al.
\newblock Cosyvoice 2: Scalable streaming speech synthesis with large language models.
\newblock \emph{arXiv preprint arXiv:2412.10117}, 2024.

\bibitem[Eskimez et~al.(2024)Eskimez, Wang, Thakker, Li, Tsai, Xiao, Yang, Zhu, Tang, Tan, et~al.]{eskimez2024e2}
Eskimez, S.~E., Wang, X., Thakker, M., Li, C., Tsai, C.-H., Xiao, Z., Yang, H., Zhu, Z., Tang, M., Tan, X., et~al.
\newblock E2 tts: Embarrassingly easy fully non-autoregressive zero-shot tts.
\newblock In \emph{2024 IEEE Spoken Language Technology Workshop (SLT)}. IEEE, 2024.

\bibitem[Fonseca \& Cohen(2024)Fonseca and Cohen]{fonseca2024can}
Fonseca, M. and Cohen, S.~B.
\newblock Can large language model summarizers adapt to diverse scientific communication goals?
\newblock \emph{arXiv preprint arXiv:2401.10415}, 2024.

\bibitem[Ho \& Salimans(2021)Ho and Salimans]{ho2021classifier}
Ho, J. and Salimans, T.
\newblock Classifier-free diffusion guidance.
\newblock In \emph{NeurIPS 2021 Workshop on Deep Generative Models and Downstream Applications}, 2021.

\bibitem[Hussain et~al.(2023)Hussain, Neekhara, Huang, Li, and Ginsburg]{hussain2023ace}
Hussain, S., Neekhara, P., Huang, J., Li, J., and Ginsburg, B.
\newblock Ace-vc: Adaptive and controllable voice conversion using explicitly disentangled self-supervised speech representations.
\newblock In \emph{ICASSP}, 2023.

\bibitem[Koluguri et~al.(2022{\natexlab{a}})Koluguri, Park, and Ginsburg]{koluguri2022titanet}
Koluguri, N.~R., Park, T., and Ginsburg, B.
\newblock Titanet: Neural model for speaker representation with 1d depth-wise separable convolutions and global context.
\newblock In \emph{ICASSP 2022-2022 IEEE International Conference on Acoustics, Speech and Signal Processing (ICASSP)}. IEEE, 2022{\natexlab{a}}.
\newblock URL \url{https://catalog.ngc.nvidia.com/orgs/nvidia/teams/nemo/models/titanet_small}.

\bibitem[Koluguri et~al.(2022{\natexlab{b}})Koluguri, Park, and Ginsburg]{koluguri2022titanetlarge}
Koluguri, N.~R., Park, T., and Ginsburg, B.
\newblock Titanet: Neural model for speaker representation with 1d depth-wise separable convolutions and global context.
\newblock In \emph{ICASSP 2022-2022 IEEE International Conference on Acoustics, Speech and Signal Processing (ICASSP)}. IEEE, 2022{\natexlab{b}}.
\newblock URL \url{https://huggingface.co/nvidia/speakerverification_en_titanet_large}.

\bibitem[Kumar et~al.(2023)Kumar, Tan, Ni, Manocha, Zhang, Henderson, and Xu]{kumar2023torchaudio}
Kumar, A., Tan, K., Ni, Z., Manocha, P., Zhang, X., Henderson, E., and Xu, B.
\newblock Torchaudio-squim: Reference-less speech quality and intelligibility measures in torchaudio.
\newblock In \emph{ICASSP 2023-2023 IEEE International Conference on Acoustics, Speech and Signal Processing (ICASSP)}, pp.\  1--5. IEEE, 2023.

\bibitem[Kumar et~al.(2024)Kumar, Seetharaman, Luebs, Kumar, and Kumar]{dac_kumar2024high}
Kumar, R., Seetharaman, P., Luebs, A., Kumar, I., and Kumar, K.
\newblock High-fidelity audio compression with improved rvqgan.
\newblock \emph{Advances in Neural Information Processing Systems}, 2024.

\bibitem[Langman et~al.(2024)Langman, Juki{\'c}, Dhawan, Koluguri, and Ginsburg]{langman2024spectral}
Langman, R., Juki{\'c}, A., Dhawan, K., Koluguri, N.~R., and Ginsburg, B.
\newblock Spectral codecs: Spectrogram-based audio codecs for high quality speech synthesis.
\newblock \emph{arXiv:2406.05298}, 2024.

\bibitem[Le et~al.(2024)Le, Vyas, Shi, Karrer, Sari, Moritz, Williamson, Manohar, Adi, Mahadeokar, et~al.]{le2024voicebox}
Le, M., Vyas, A., Shi, B., Karrer, B., Sari, L., Moritz, R., Williamson, M., Manohar, V., Adi, Y., Mahadeokar, J., et~al.
\newblock Voicebox: Text-guided multilingual universal speech generation at scale.
\newblock \emph{Advances in Neural Information Processing Systems}, 2024.

\bibitem[Li et~al.(2024)Li, Han, Raghavan, Mischler, and Mesgarani]{li2024styletts}
Li, Y.~A., Han, C., Raghavan, V., Mischler, G., and Mesgarani, N.
\newblock Styletts 2: Towards human-level text-to-speech through style diffusion and adversarial training with large speech language models.
\newblock \emph{Advances in Neural Information Processing Systems}, 2024.

\bibitem[Mentzer et~al.(2024)Mentzer, Minnen, Agustsson, and Tschannen]{mentzer2024finite}
Mentzer, F., Minnen, D., Agustsson, E., and Tschannen, M.
\newblock Finite scalar quantization: {VQ}-{VAE} made simple.
\newblock In \emph{The Twelfth International Conference on Learning Representations}, 2024.

\bibitem[Neekhara et~al.(2024{\natexlab{a}})Neekhara, Hussain, Ghosh, Li, Valle, Badlani, and Ginsburg]{t5tts}
Neekhara, P., Hussain, S., Ghosh, S., Li, J., Valle, R., Badlani, R., and Ginsburg, B.
\newblock Improving robustness of llm-based speech synthesis by learning monotonic alignment.
\newblock \emph{INTERSPEECH}, 2024{\natexlab{a}}.

\bibitem[Neekhara et~al.(2024{\natexlab{b}})Neekhara, Hussain, Valle, Ginsburg, Ranjan, Dubnov, Koushanfar, and Mcauley]{neekharaselfvc}
Neekhara, P., Hussain, S.~S., Valle, R., Ginsburg, B., Ranjan, R., Dubnov, S., Koushanfar, F., and Mcauley, J.
\newblock {S}elf{VC}: Voice conversion with iterative refinement using self transformations.
\newblock In \emph{Proceedings of the 41st International Conference on Machine Learning}, 2024{\natexlab{b}}.

\bibitem[Oliveira et~al.(2023)Oliveira, Casanova, Junior, Soares, and Galv{\~a}o~Filho]{oliveira2023cml}
Oliveira, F.~S., Casanova, E., Junior, A.~C., Soares, A.~S., and Galv{\~a}o~Filho, A.~R.
\newblock {CML-TTS}: A multilingual dataset for speech synthesis in low-resource languages.
\newblock In \emph{International Conference on Text, Speech, and Dialogue}, pp.\  188--199. Springer, 2023.

\bibitem[Ouyang et~al.(2022)Ouyang, Wu, Jiang, Almeida, Wainwright, Mishkin, Zhang, Agarwal, Slama, Ray, et~al.]{ouyang2022rlhf}
Ouyang, L., Wu, J., Jiang, X., Almeida, D., Wainwright, C., Mishkin, P., Zhang, C., Agarwal, S., Slama, K., Ray, A., et~al.
\newblock Training language models to follow instructions with human feedback.
\newblock \emph{Advances in Neural Information Processing Systems}, 2022.

\bibitem[Pratap et~al.(2020)Pratap, Xu, Sriram, Synnaeve, and Collobert]{pratap20_interspeech}
Pratap, V., Xu, Q., Sriram, A., Synnaeve, G., and Collobert, R.
\newblock {MLS: A Large-Scale Multilingual Dataset for Speech Research}.
\newblock In \emph{INTERSPEECH}, 2020.

\bibitem[Radford et~al.(2022)Radford, Kim, Xu, Brockman, McLeavey, and Sutskever]{radford2022whisper}
Radford, A., Kim, J.~W., Xu, T., Brockman, G., McLeavey, C., and Sutskever, I.
\newblock Robust speech recognition via large-scale weak supervision, 2022.

\bibitem[Rafailov et~al.(2024)Rafailov, Sharma, Mitchell, Manning, Ermon, and Finn]{rafailov2024direct}
Rafailov, R., Sharma, A., Mitchell, E., Manning, C.~D., Ermon, S., and Finn, C.
\newblock Direct preference optimization: Your language model is secretly a reward model.
\newblock \emph{Advances in Neural Information Processing Systems}, 2024.

\bibitem[Sahoo et~al.(2024)Sahoo, Meharia, Ghosh, Saha, Jain, and Chadha]{sahoo2024comprehensive}
Sahoo, P., Meharia, P., Ghosh, A., Saha, S., Jain, V., and Chadha, A.
\newblock A comprehensive survey of hallucination in large language, image, video and audio foundation models.
\newblock \emph{Findings of the Association for Computational Linguistics: EMNLP}, 2024.

\bibitem[Sanchez et~al.(2023)Sanchez, Fan, Spangher, Levi, Ammanamanchi, and Biderman]{sanchez2023stay}
Sanchez, G., Fan, H., Spangher, A., Levi, E., Ammanamanchi, P.~S., and Biderman, S.
\newblock Stay on topic with classifier-free guidance.
\newblock \emph{arXiv preprint arXiv:2306.17806}, 2023.

\bibitem[Shao et~al.(2024)Shao, Wang, Zhu, Xu, Song, Bi, Zhang, Zhang, Li, Wu, et~al.]{shao2024deepseekmath}
Shao, Z., Wang, P., Zhu, Q., Xu, R., Song, J., Bi, X., Zhang, H., Zhang, M., Li, Y., Wu, Y., et~al.
\newblock Deepseekmath: Pushing the limits of mathematical reasoning in open language models.
\newblock \emph{arXiv:2402.03300}, 2024.

\bibitem[Smirnov(2024)]{smirnov2024classifier}
Smirnov, R.
\newblock Classifier-free guidance in llms safety.
\newblock \emph{arXiv preprint arXiv:2412.06846}, 2024.

\bibitem[Song et~al.(2024)Song, Chen, Wang, Ma, and Chen]{song2024ella}
Song, Y., Chen, Z., Wang, X., Ma, Z., and Chen, X.
\newblock Ella-v: Stable neural codec language modeling with alignment-guided sequence reordering.
\newblock \emph{arXiv preprint arXiv:2401.07333}, 2024.

\bibitem[Touvron et~al.(2023)Touvron, Lavril, Izacard, Martinet, Lachaux, Lacroix, Rozi{\`e}re, Goyal, Hambro, Azhar, et~al.]{touvron2023llama}
Touvron, H., Lavril, T., Izacard, G., Martinet, X., Lachaux, M.-A., Lacroix, T., Rozi{\`e}re, B., Goyal, N., Hambro, E., Azhar, F., et~al.
\newblock Llama: Open and efficient foundation language models.
\newblock \emph{arXiv preprint arXiv:2302.13971}, 2023.

\bibitem[Wang et~al.(2023)Wang, Chen, Wu, Zhang, Zhou, Liu, Chen, Liu, Wang, Li, et~al.]{wang2023neural}
Wang, C., Chen, S., Wu, Y., Zhang, Z., Zhou, L., Liu, S., Chen, Z., Liu, Y., Wang, H., Li, J., et~al.
\newblock Neural codec language models are zero-shot text to speech synthesizers.
\newblock \emph{arXiv:2301.02111}, 2023.

\bibitem[Wang et~al.(2024)Wang, Thakker, Chen, Kanda, Eskimez, Chen, Tang, Liu, Li, and Yoshioka]{wang2024speechx}
Wang, X., Thakker, M., Chen, Z., Kanda, N., Eskimez, S.~E., Chen, S., Tang, M., Liu, S., Li, J., and Yoshioka, T.
\newblock Speech{X}: Neural codec language model as a versatile speech transformer.
\newblock \emph{IEEE/ACM Transactions on Audio, Speech, and Language Processing}, 2024.

\bibitem[Xu et~al.(2023)Xu, Jia, Majumdar, Huang, Watanabe, and Ginsburg]{xu2023efficient}
Xu, H., Jia, F., Majumdar, S., Huang, H., Watanabe, S., and Ginsburg, B.
\newblock Efficient sequence transduction by jointly predicting tokens and durations.
\newblock In \emph{International Conference on Machine Learning}. PMLR, 2023.
\newblock URL \url{https://huggingface.co/nvidia/parakeet-tdt-1.1b}.

\bibitem[Yang et~al.(2024)Yang, Tian, Tan, Huang, Liu, Guo, Chang, Shi, Bian, Zhao, et~al.]{yanguniaudio}
Yang, D., Tian, J., Tan, X., Huang, R., Liu, S., Guo, H., Chang, X., Shi, J., Bian, J., Zhao, Z., et~al.
\newblock Uniaudio: Towards universal audio generation with large language models.
\newblock In \emph{Forty-first International Conference on Machine Learning}, 2024.

\bibitem[Zeghidour et~al.(2021)Zeghidour, Luebs, Omran, Skoglund, and Tagliasacchi]{zeghidour2021soundstream}
Zeghidour, N., Luebs, A., Omran, A., Skoglund, J., and Tagliasacchi, M.
\newblock Sound{S}tream: An end-to-end neural audio codec.
\newblock \emph{IEEE/ACM Transactions on Audio, Speech, and Language Processing}, 2021.

\bibitem[Zen et~al.(2019)Zen, Dang, Clark, Zhang, Weiss, Jia, Chen, and Wu]{zen2019libritts}
Zen, H., Dang, V., Clark, R., Zhang, Y., Weiss, R.~J., Jia, Y., Chen, Z., and Wu, Y.
\newblock Libri{TTS}: A corpus derived from librispeech for text-to-speech.
\newblock \emph{INTERSPEECH}, 2019.

\bibitem[Zhang et~al.(2024)Zhang, Li, Li, Zhang, Wang, Zhou, and Qiu]{zhang2024speechalign}
Zhang, D., Li, Z., Li, S., Zhang, X., Wang, P., Zhou, Y., and Qiu, X.
\newblock Speechalign: Aligning speech generation to human preferences.
\newblock In \emph{Adances in Neural Information Processing Systems}, 2024.

\bibitem[Zhang et~al.(2023)Zhang, Zhou, Wang, Chen, Wu, Liu, Chen, Liu, Wang, Li, et~al.]{zhang2023speak}
Zhang, Z., Zhou, L., Wang, C., Chen, S., Wu, Y., Liu, S., Chen, Z., Liu, Y., Wang, H., Li, J., et~al.
\newblock Speak foreign languages with your own voice: Cross-lingual neural codec language modeling.
\newblock \emph{arXiv:2303.03926}, 2023.

\end{thebibliography}
